\documentclass[a4paper,11pt]{article}
\usepackage{jheppub}
\usepackage{mathtools}
\usepackage{bigints}
\usepackage{braket}
\usepackage{tikz}
\usetikzlibrary{decorations.pathmorphing,decorations.markings,arrows.meta}
\allowdisplaybreaks
\title{The Wilson-line-dressed charged sector of scalar QED: superselection and the infraparticle}
\author[a]{Gordon W. Semenoff}
\author[a]{and Conor Waterfield}

\affiliation[a]{Department of Physics and Astronomy, University of British Columbia,\\
6224 Agricultural Road, Vancouver, British Columbia, Canada V6T 1Z1}

\emailAdd{gordonws@phas.ubc.ca}
\abstract{We develop a manifestly gauge and Lorentz invariant perturbative calculus to study the finite time and distance evolution of the electrically charged states created by Wilson-line-dressed scalar field operators in scalar quantum electrodynamics.  By an Osterwalder-Schrader reflection and Euclidean gluing, the overlap of two dressed states is expressed as a Euclidean 2 point function of dressed operators, which is then evaluated in renormalized perturbation theory with a Stueckelberg photon mass as a gauge, BRST and Lorentz invariant infrared cutoff. This renders explicit and computable in closed form several structural features of the charged sector that have so far been accessible mainly through non-perturbative or algebraic arguments. We find that states dressed by non-parallel Wilson lines are ``cloud orthogonal". Their overlap vanishes as a power of the infrared regulator, with an exponent that depends only on the angle between the two dressings and is an infrared analogue of the cusp anomalous dimension with the difference that it is one-loop exact whenever the charged matter is mass-gapped.  Charged states are thereby superselected by the orientation of their dressing. Within a superselection sector the 2 point function is infrared finite but exhibits the infraparticle. The mass shell pole is replaced by the edge of a branch cut. As a result, the asymptotic spread of a dressed charge is a power law in proper time with a  one-loop-exact computable  exponent depending on the dressing and on the direction of motion. We show that although cloud orthogonality and the infraparticle are each extremely sensitive to fundamental infrared cutoffs, the infraparticle scaling law persists across a wide window of proper time.}

\keywords{Gauge Symmetry, Scattering Amplitudes, Nonperturbative Effects}   

\begin{document}
\maketitle

\section{Introduction and Summary}
\label{introduction}
It is by now apparent that physical states of charged particles in an
Abelian gauge theory such as quantum electrodynamics must be dressed states.
This is required for the existence of the S matrix, where  applying the
Faddeev--Kulish dressing \cite{Kulish:1970ut} to incoming and outgoing asymptotic charged particles
cancels the infrared
divergences which occur in the perturbative computation of S matrix
elements, rendering transition amplitudes infrared finite
\cite{Chung:1965zza,Kibble:1969ip}.  It resolves subtle information theoretic issues
\cite{Carney:2017jut,Carney:2017oxp,Carney:2018ygh}
which arise in the Bloch-Nordsieck approach \cite{BlochNordsieck,YFS} to the infrared problem.  The latter computes transition probabilities for undressed particles inclusive of soft photon production. (Weinberg's book \cite{weinberg1995quantum1} contains a particularly clear summary of that approach.)  Soft photon dressing is also needed for the S matrix to satisfy selection rules that arise from
 asymptotic gauge symmetries \cite{Strominger:2017zoo}.
 
Moreover, it is by now understood that the need for dressing is more than a simple artifact of demanding an S matrix in a theory which shouldn't have one.  It has an immediately obvious symptom in the structure of quantum states, the fact that quantum electrodynamics does not have a local, gauge invariant operator which creates a charged particle \cite{Ruggero}.  

In this paper we will study a class of nonlocal gauge invariant electrically charged operators which are the complex scalar fields of scalar quantum electrodynamics dressed by semi-infinite Wilson lines,
\begin{equation}
\begin{aligned}\label{dress}
&
\Phi_v^*(x)~=~\phi^*(x)~\mathcal P e^{ ie\int_{-\infty}^0 v^\mu ds A_\mu(x+sv)},~\Phi_{v}(x)~=~\Phi_v^{*\dagger}(x)=\bar{\mathcal P}~e^{ -ie\int_{-\infty}^0 v^\mu ds A_\mu(x+sv)}\phi(x)
\end{aligned}
\end{equation}
where the symbol $\mathcal P$ ($\bar{\mathcal P}$) indicates that the Wilson line is path-ordered (anti-path-ordered) and $v^\mu$ is a constant 4-vector, which we will take to be time-like and forward directed, although the Euclidean computation itself does not depend on this choice.\footnote{The Euclidean correlator which we shall find in equation (\ref{Dv}) is defined
for any Euclidean direction $v^\mu$ and may be continued to Minkowski
dressings with either timelike or spacelike $v^\mu$.    We
restrict attention to the timelike case, in which the dressing is the
Coulomb cloud of a charge of velocity $v^\mu$.  A spacelike Wilson line
instead creates a flux tube of extensive energy, Mandelstam's string
\cite{Mandelstam:1962mi}, and states carrying non-coincident strings are
orthogonal for reasons unrelated to the infrared exponents computed here. 
A controlled treatment requires the string-localized fields of references
\cite{MSY,MSY1,MRS}, and we shall not pursue it here.}
We will use the leading orders of renormalized perturbation theory with some resummation arguments.  We will cut off the infrared singularities by using the St\"uckelberg mechanism to introduce a gauge invariant photon mass. 
This is also a Lorentz invariant infrared cutoff.

The idea of dressing a charged local operator with a coherent state-like cloud of photons in such a way that the resulting operator is gauge invariant
goes back to Jordan \cite{Jordan}, Dirac \cite{Dirac:1955uv} and Mandelstam \cite{Mandelstam:1962mi}.
Dirac  proposed a dressing which is the exponential of a linear functional of the photon field,
 \begin{equation}\label{dirac dressing}
 \begin{aligned}
&
\Phi^*_\xi(x)~=~\phi^*(x) ~e^{ ie\int d^4y \xi^{\mu}(x-y) A_\mu(y)}~~,~~ 
\Phi_\xi(x)~=~e^{-ie\int d^4y \xi^{*\mu}(x-y) A_\mu(y)}~\phi(x)
\\&
\partial_\mu \xi^\mu(x-y)=\delta^{(4)}(x-y)
\end{aligned}
\end{equation}
The semi-infinite Wilson line that we use (\ref{dress}) is a specific example which has 
the kernel $\xi^\mu(x-y)$ concentrated along the line. What is essential in such a dressing is
the asymptotic behaviour which must obey (written in Euclidean signature)
\begin{equation}
\lim_{|x|\to \infty}\int_{S^3}d\Omega |x|^3 \hat x\cdot\xi(|x|\hat x-y)=1
\end{equation}
or, for the Fourier transform
\begin{equation}
\xi^\mu(k) = -i\frac{v^\mu}{k_\nu v^\nu}+ {\rm~less~singular~at~small~}k
\end{equation}
The essential part of $\xi^\mu(x-y)$ is its asymptotic behaviour which is encoded in  a 4-vector $v^\mu$.
It coincides with the constant vector  in our semi-infinite Wilson lines, where the kernel would be
\begin{equation}
\begin{aligned}
&\xi^\mu_{WL}(x-y) =\int_{-\infty}^0ds ~v^\mu~\delta^4(x-y+sv)
\\&\xi^\mu_{WL}(k)= \int_{-\infty}^0 ds~v^\mu~ e^{isv\cdot k} = -i\frac{v^\mu}{k\cdot v-i\varepsilon}
\end{aligned}
\end{equation}

Although the operators that we are discussing here have a priori nothing to do with asymptotic states, one might notice the formal similarity of this dressing with the Faddeev-Kulish dressing of an asymptotic state where the in-state of a charged particle  with momentum $p$ is replaced by 
\begin{equation}\label{FK}
c^\dagger_{\rm in}(p)\ket{0}~\longrightarrow~
\exp\left(e\int \frac{d^3k}{(2\pi)^32|\vec k|} \frac{p_\mu}{p\cdot k}[a^\mu_{\rm in} (k)-a^{\dagger\mu}_{\rm in}(k)]\right)c^\dagger_{\rm in}(p)\ket{0}
\end{equation} 
where $a^\mu_{\rm in}(k),a^{\dagger\mu}_{\rm in}(k)$ are in-state photon annihilation and creation operators. In that case, the analog of $\xi^\mu$ is $\xi^\mu(k)=-i\frac{p^\mu}{p\cdot k}$ and the 4-vector is $v^\mu=\frac{1}{m}p^\mu$ which is the 4-velocity of the charged particle.

Rather than asymptotic states or the S matrix, in the following we will  focus on evolution over finite times and distances of dressed quantum states with electric charge.  We will address the
following question.
 Let us say that we create a gauge invariant charged state in the vicinity of a spacetime event by operating with a suitably smeared Wilson-line-dressed charged  operator on the vacuum to produce, for example,
\begin{equation}\label{state 0}
\ket{f_0,v}~\equiv~\int d^4y f_0(y)\Phi_v^*(y)\ket{0}
\end{equation}
where $\Phi_v^*(y)$ is the Wilson-line-dressed operator from equation (\ref{dress}) and 
the envelope function $f_0(y)$ is needed to normalize the state.  It is localized near the event $y^\mu=0$.
We can regard the Wilson line dressing as a record of how this state was prepared.  It is the result of parallel transport of the charged field along a flat space-time geodesic (straight line) stretching from a point at infinity to a point where the charged particle is created (the parametric equation of the line is  $x^\mu(s)=v^\mu s +y^\mu$, $s\in(-\infty,0)$). 

Then we ask for the probability amplitude that the state in (\ref{state 0}) evolves to another state which is similar but is localized in the vicinity of 
another event $x^\mu$ with position $\vec x$ after a time $x^0$, and defined with dressing 4-velocity ${v'}^\mu$, that is, to the state
\begin{equation}\label{state x}
\ket{f_x,v'}~\equiv~\int d^4y f_x(y)\Phi_{v'}^*(y)\ket{0}
\end{equation}
with $f_x(y)$ concentrated near $y^\mu \sim x^\mu$.  The quantum amplitude for this evolution is the inner product of the two states, $\braket{f_x,v'|f_0,v}$.  \footnote{
Throughout, we shall study the inner product, which is the Wightman function
of the dressed fields and, for positive times, coincides with the
time-ordered two-point function in which each dressed operator is ordered as
a unit, the ordering produced by the in-in formalism; it is not the naive
time-ordered product of all of the constituent fields, which would describe
a charge and an anti-charge present for all past time.
Moreover, questions about
the propagation of disturbances in a quantum field theory are
properly answered  by retarded response functions. All of the two-point functions
of a given pair of operators are determined by the spectral function of their commutator and
they differ only in the kernel against which  the spectral function is integrated.  The cloud orthogonality
(\ref{orthogonality}) which we shall find is the vanishing of the mixed
spectral function as $m_\gamma\to0$, and the infraparticle scaling law
(\ref{infraparicle scaling}) is a consequence of the threshold behaviour of
the diagonal spectral function.  The retarded response function of dressed
fields therefore exhibits the same behaviour.}

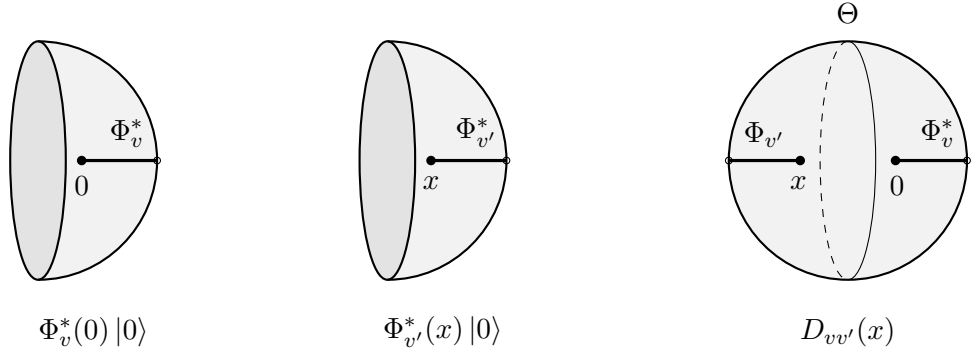
\begin{figure}[h]
\centering
\begin{tikzpicture}[scale=1.05]
\begin{scope}[shift={(0,0)}]
  \fill[gray!10] (0,-1.5) arc (-90:90:1.5) arc (90:270:0.35 and 1.5) -- cycle; 
  \fill[gray!22] (0,0) ellipse (0.35 and 1.5);      
  \draw[thick] (0,-1.5) arc (-90:90:1.5);           
  \draw[thick] (0,0) ellipse (0.35 and 1.5);        
  \draw[very thick] (0.55,0) -- (1.5,0);            
  \fill (0.55,0) circle (1.7pt);
  \node[below=1pt] at (0.55,-0.04) {$0$};
  \node[above] at (1.12,0.04) {$\Phi^*_{v}$};
  \draw (1.5,0) circle (1.2pt);
  \node at (0.7,-2.15) {$\Phi_{v}^*(0)\ket{0}$};
\end{scope}
\begin{scope}[shift={(4.4,0)}]
  \fill[gray!10] (0,-1.5) arc (-90:90:1.5) arc (90:270:0.35 and 1.5) -- cycle;
  \fill[gray!22] (0,0) ellipse (0.35 and 1.5);
  \draw[thick] (0,-1.5) arc (-90:90:1.5);
  \draw[thick] (0,0) ellipse (0.35 and 1.5);
  \draw[very thick] (0.55,0) -- (1.5,0);
  \fill (0.55,0) circle (1.7pt);
  \node[below=1pt] at (0.55,-0.04) {$x$};
  \node[above] at (1.12,0.04) {$\Phi^*_{v'}$};
  \draw (1.5,0) circle (1.2pt);
  \node at (0.7,-2.15) {$\Phi^*_{v'}(x)\ket{0}$};
\end{scope}
\begin{scope}[shift={(10.2,0)}]
  \fill[gray!10] (0,0) circle (1.5);
  \draw[thick] (0,0) circle (1.5);
  \draw[dashed] (0,1.5) arc (90:270:0.35 and 1.5);  
  \draw (0,-1.5) arc (-90:90:0.35 and 1.5);         
  \node[above] at (0,1.6) {$\Theta$};
  \draw[very thick] (0.6,0) -- (1.5,0);
  \fill (0.6,0) circle (1.7pt);
  \node[below=1pt] at (0.62,-0.04) {$0$};
  \node[above] at (1.13,0.04) {$\Phi^*_{v}$};
  \draw (1.5,0) circle (1.2pt);
  \draw[very thick] (-0.6,0) -- (-1.5,0);
  \fill (-0.6,0) circle (1.7pt);
  \node[below=1pt] at (-0.62,-0.04) {$x$};
  \node[above] at (-1.05,0.04) {$\Phi_{v'}$};
  \draw (-1.5,0) circle (1.2pt);
  \node at (0,-2.15) {$D_{vv'}(x)$};
\end{scope}
\end{tikzpicture}
\caption{The Euclidean gluing interpretation of the dressed correlator:  The left and
centre diagrams represent wave-functionals, at a Euclidean time within the interval $[0,x^0]$, corresponding to the states $\Phi_{v}^*(0)\ket{0}$ and $\Phi^*_{v'}(x)\ket{0}$ created by the dressed operator insertions.   The  wave-functionals of those states are
obtained by taking the functional integral over fields living in the half-spaces with the functions obeying
prescribed Dirichlet boundary conditions at the half-space boundaries, the boundaries being the surfaces on the left of each figure.
The operators
$\Phi_{v}^*$ and $\Phi_{v'}^*$ are inserted into the functional integrals at positions $0$ and $x$, respectively, with the Wilson lines running
to the boundary at infinity, along the contours $y^\mu(s)=v^\mu s$ and $y^\mu(s)=x^\mu+{v'}^\mu s$, respectively, where $v^\mu$ and ${v'}^\mu$ are now Euclidean unit vectors.
In the right-hand-diagram the inner product of the states
$\Phi_{v}^*(0)\ket{0}$ and $\Phi^*_{v'}(x)\ket{0}$ is
computed by sewing the half-spaces together along their boundaries -- identifying the boundary values of the fields and functionally integrating over them -- after the
Osterwalder--Schrader reflection \cite{OS}, $\Theta$, of the state $\Phi_{v'}^*(x)\ket{0}$. This obtains the Euclidean 2 point function $D_{vv'}(x)$ in equation (\ref{Dv}). The quantum  amplitude is obtained from this Euclidean correlator by inverse Wick rotation and analytic continuation of $v,v'$.}
\label{euclidean inner product}
\end{figure}

A perturbative computation of the inner product $\braket{f_x,v'|f_0,v}$ of the state in equation (\ref{state x}) and the state in equation (\ref{state 0}) would require the in-in formalism of quantum field theory.  
Alternatively, it is technically easier to use the Euclidean time functional integral gluing prescription to compute the inner products of the states in Euclidean time, as is described in figure \ref{euclidean inner product} and subsequently inverse Wick rotate back to Lorentzian signature time. The result  is that the  inner product is obtained from the appropriate inverse Wick rotation of the Euclidean space 2 point function
\begin{equation}\label{Dv}
D_{vv'}(x)~\equiv ~\bigg<~e^{ ie\int_0^{\infty}ds~{v'}^\mu A_\mu(x+sv')}~ \phi(x)~\phi^*(0)~
e^{ie\int_{-\infty}^0 ds\, v^\mu A_\mu(sv)}~ \bigg>
\end{equation}
(where $x^\mu,0^\mu,v^\mu,{v'}^\mu$ are now Euclidean 4-vectors)
The bracket  on the right-hand-side of (\ref{Dv}) is computed by a Euclidean 
functional integral, some details of which are given in section \ref{model}.  Within this bracket,
the conjugated Wilson line is time-reversed.  The time reverse is due to the Osterwalder-Schrader reflection \cite{OS} which must operate on the integration variables in the dual state in the inner product when it is computed in Euclidean space, as depicted in figure \ref{euclidean inner product}.

Once we have reduced the problem to the computation of a 2 point function, studying it in the leading orders of quantum electrodynamics perturbation theory is straightforward and elementary.  
To the extent that we compute, we confirm that the 2 point function is ultraviolet renormalizable with the same counterterms as would be used to make the S matrix finite in native scalar QED plus a multiplicative renormalization of the composite dressed operators 
$$
\Phi_{v}(x),~\Phi_{v}^*(x)~\longrightarrow~ z^\frac{1}{2}\Phi_{v}(x),~z^\frac{1}{2}\Phi^*_{v}(x)
$$ 

 It has been known for a very long time in the axiomatic  algebraic approach to quantum electrodynamics that   the  
long-ranged photon clouds that are needed in order for an electrically charged quantum state to satisfy Gauss' law, and thereby be gauge invariant,   make the quantum states with significantly differing clouds orthogonal to each other \cite{Schroer:1963,Frohlich:1979nn,Frohlich:1979is,Buchholz:1982ed} (see also \cite{Haag} Chap.VI for a comprehensive review).  They live in different Hilbert spaces.   The same effect  leads to spontaneous breaking of the Lorentz symmetry for a given cloud 
\cite{Frohlich:1979nn,Frohlich:1979is,Buchholz:1982ed,Balachandran:2013wca,Balachandran:2014nea,Campiglia:2025ssb}.

Our perturbative computation of the Euclidean 2 point function of dressed operators confirms  this ``cloud orthogonality''.
We shall show that 
 the entirety of the infrared divergent parts can be assembled into an overall factor which, in the limit where the photon mass $m_\gamma$ approaches zero, appears in the Euclidean 2 point function in the form 
\begin{equation}\label{cusp 1}
D_{vv'}(p)~=~\left(\frac{m_\gamma}{m}\right)^{\frac{e^2}{4\pi^2}(\delta\cot\delta-1)}     \bar D_{vv'}(p)
~,~~~\cos\delta~ \equiv~\frac{v\cdot v'}{|v||v'|}
\end{equation}
where $m$ is  the mass of the charged scalar field,   $\delta$ is the angle between the Euclidean vectors $v$ and $v'$ and the coefficient function $\bar D_{vv'}(p)$ is finite and cutoff independent as $m_\gamma\to 0$.  

The Euclidean 2 point function in (\ref{cusp 1}) diverges when $\delta\neq 0$ and as $m_\gamma\to 0$. The infrared cutoff dependent coefficient in (\ref{cusp 1}) is a function of $v$ and $v'$ which does not factorize into a function of $v$ and a function of $v'$.
It therefore cannot be removed by multiplicative renormalization of the composite operators  $\Phi_{v'}(x)$ and $\Phi_{v}^*(x)$ where the renormalization constants would each have to depend on their $v$ alone.

Equation (\ref{cusp 1}) is the behaviour of the Euclidean space correlation function. To find the inner product of the states $\ket{f_x,v'}$ and $\ket{f_0,v}$, we must inverse Wick rotate to Minkowski space.  Under that rotation, the angle between the velocity vectors of the Wilson lines, $\delta$,  is replaced by $-i\chi$ where $\chi$ is the relative rapidity of the dressing vectors (assuming that they are both forward-directed, timelike vectors), $v^\mu {v_\mu}'=-\cosh\chi$ in Minkowski space when $v^\mu v_\mu=-1={v'}^\mu{v'}_\mu$, and the factor in equation (\ref{cusp 1}) becomes
\begin{equation}\label{factor}
\begin{aligned}
&\braket{f_x,v'|f_0,v} ~=~\left(\frac{m_\gamma}{m}\right)^{\frac{e^2}{4\pi^2}(\chi\coth\chi -1)} \times 
\biggl[ {\rm~nonzero~and~finite~as~}m_\gamma\to 0~ \biggr]
\\&v^\mu v_\mu'=-\cosh\chi
\end{aligned}
\end{equation}
In this expression, the exponent is positive semi-definite and vanishes only when $\chi=0$. As a consequence, once the infrared regulator is removed, that is, $m_\gamma\to0$, states that are created by dressed operators with non-parallel $v$'s are indeed orthogonal to each other, 
\begin{align}\label{orthogonality}
\lim_{m_\gamma\to0} ~\braket{f_x,v'|f_0,v} =0~,~~\forall~v\neq v'
\end{align}

If we implement our thought experiment where we create the (suitably smeared) state $\int d^4y f(y)\Phi_v^*(y)\ket{0}$, the amplitude for it to evolve to, or to have evolved from, any state $\int d^4y \tilde f(y)\Phi_{v'}^*(y)\ket{0}$ with $v'\neq v$ is zero.  

This implies a superselection rule for dressed charged operators with different values of $v$. They create states which live in mutually orthogonal Hilbert spaces. In as much as the Wilson line orientation records the history of how the charged state was prepared, states turn out to be orthogonal unless their histories have identical $v$'s.

   This superselection is closely related to, and can be regarded as the finite time signature of the decoherence phenomenon for undressed asymptotic Fock states noted  by Carney et al.~\cite{Carney:2017jut}.  In that case, decoherence suppresses the scattering of wave-packets of any but correctly dressed Faddeev-Kulish asymptotic charged particle states \cite{Carney:2018ygh}. 
   
The exponent in the factor in (\ref{factor}) is formally similar to the Abelian cusp anomalous dimension \cite{Polyakov:1980ca} of cusped Wilson lines in the one-loop approximation.  It is geometrically similar to the cusp dimension because the very long wave-length photons that contribute to the infrared divergences do not distinguish the geometry of the semi-infinite Wilson lines from similar lines which would  intersect at a cusp.  On the other hand, it differs from the ultraviolet cusp anomalous dimension in that, after the exponentiation of one loop order to get the exponential in (\ref{factor}), it is one-loop exact.  This is a consequence of the soft photon theorems \cite{Low:1954kd,Low:1958sn,Burnett:1967km,Gribov:1966hs}, which render the coupling of infrared photons to the mass gapped charged fields eikonal and their effective action Gaussian, together with the infrared safety of the vacuum polarization.  The ultraviolet cusp gets corrected at all orders, where it is known to four loops \cite{bruser}.

The reader should beware that the orthogonality (\ref{orthogonality}) resulting from (\ref{cusp 1}) is an idealization.  Any relaxation of the limits of having exactly massless photons travelling in open, infinite spacetime could make it  significantly less dramatic.  See the last paragraphs of section \ref{model} for a discussion of this fact. We  will also have more to say about it here, shortly.

Closely related to cloud orthogonality is a phenomenon which Schroer called the ``infraparticle'' \cite{Schroer:1963}.  The gauge invariant charged sector states contain superpositions of momentum eigenstates which are then also gauge invariant eigenstates of the mass operator $-P_\mu P^\mu$.  In the infraparticle scenario, the entire spectrum of $-P_\mu P^\mu$ begins at the charged particle mass squared and it is continuous.  There is no discrete eigenstate, which would be the mathematical signal of the presence of an isolated single charge particle state. 
Instead, an enhanced continuum spectrum appears near the charged particle mass $m^2$  called the ``infraparticle''.   Buchholz \cite{Buch} and Fr\"ohlich et al. \cite{Frohlich:1979nn} tied the infraparticle to the spontaneous breaking of Lorentz symmetry by showing that it could not simultaneously be gauge invariant and carry a unitary representation of the Lorentz transformations. 

What we shall find is also consistent with this picture.  When we respect the cloud superselection rule and study the 2 point function with parallel Wilson line dressings, $D_{vv}(x)$, the logarithms of the small photon mass are absent.    However, removal of the infrared cutoff leads to mass-shell logarithms which take the form
\begin{equation}\label{infraparticle exponent 0}
\begin{aligned}
&D_{vv}(p)~=~\frac{1}{p^{2}+m^{2}}\biggr\{ 1+\# \ln\frac{p^2+m^2}{m^2}+\ldots\biggr\}
\end{aligned}
\end{equation}
The mass-shell logarithms are a vestige of the infrared divergences and their appearance is called the residue obstruction. 
They prevent the 2 point function from having an isolated pole.  They are already a well-known property of the undressed charged particle 2-point function \cite{Abrikosov:1956} which can be found discussed in textbooks \cite{peskin1995introduction}\cite{Sterman}\cite{semenoff2023quantum} and which we will review in section \ref{self energy}.  
The spectral representation  \cite{Umezawa,Kallen,Lehmann} then tells us that the intermediate states in the 2 point function are the infraparticle continuum. 

Our full result, with the infrared logarithms resummed, and inverse Wick rotated to Minkowski space, takes the form
 \begin{equation}\label{infraparticle exponent 1}
\begin{aligned}
&D_{vv}(p)~=~-i\left[\frac{1}{p^2+m^2-i\epsilon}\right]^{1-\gamma(\zeta)}
+\ldots 
\\
&\gamma(\zeta)=\frac{e^{2}}{2\pi^{2}}\bigl(\zeta\coth\zeta-1\bigr)
\\
& p_\mu v^\mu=-m\cosh\zeta
\end{aligned}
\end{equation}
where the ellipses stand for contributions which remain finite at $p^2\to -m^2$ and $\gamma(\zeta)$ is a computable exponent, given at one-loop order in the above formula and which, after the re-summation which exponentiates it as in (\ref{infraparticle exponent 1}), is one-loop exact.  It has a similar nature and origin to the cloud orthogonality exponent.  In the semi-classical limit it arises from the cusps in the junctions between the dressing Wilson lines and the classical propagation trajectory of the charged particle. It is for that reason that it contains  the same function of a different rapidity angle -- the relative rapidity of the Wilson lines and the semi-classical propagation of the particle. There are two cusps, one at each junction, thus the relative factor of two in the coefficient, $\frac{e^2}{2\pi^2}$ versus $\frac{e^2}{4\pi^2}$.  It is the exponent $\gamma(\zeta)$ which is of the greatest interest to us, as we shall discuss shortly.

In a remarkable series of papers, Bagan, Lavelle and McMullan
\cite{Bagan:1996pka,Bagan:1997zr,Bagan:1999jf} studied the 2 point function 
\begin{equation}\label{BLM 2 point function}
D_{{\rm Coul.,}v}(p)= \int d^4x e^{-ipx}\bra{0}\mathcal T \Phi_{{\rm Coul.,}v}(x)\Phi^*_{{\rm Coul.,}v}(0)\ket{0}
\end{equation}
of charged field operators dressed by boosted Coulomb fields\footnote{
Bagan, Lavelle and McMullan can avoid the in-in formalism and simply study the time-ordered 2 point function (\ref{BLM 2 point function}) because their boosted Coulomb dressings occur on fixed time slices.}  
\begin{equation}\label{v coulomb dressing}
\begin{aligned}
&\Phi_{{\rm Coul.,}v}(x)~= ~\phi(\vec x,t)~e^{
-\frac{ie\gamma}{4\pi} \int d^3y \frac{
\gamma^{-1}\nabla_1A_1(\vec y,t)+\nabla_2A_2(\vec y,t)+\nabla_3A_3(\vec y,t)-v^1E_1(\vec y,t)
}{
\sqrt{  \gamma^2(x_1-y_1)^2+(x_2-y_2)^2+(x_3-y_3)^2  }
}
}
\\&
\Phi_{{\rm Coul.,}v}^*(x)~=~\phi^*(\vec x,t)~e^{
\frac{ie\gamma}{4\pi} \int d^3y \frac{
\gamma^{-1}\nabla_1A_1(\vec y,t)+\nabla_2A_2(\vec y,t)
+\nabla_3A_3(\vec y,t)-v^1E_1(\vec y,t)
}{
\sqrt{  \gamma^2(x_1-y_1)^2+(x_2-y_2)^2+(x_3-y_3)^2  }
}
}
\\&
\vec v=(v^1,0,0),~\gamma=1/\sqrt{1-\vec v^2},~v^\mu=(\gamma, \gamma\vec v)
\end{aligned}
\end{equation}
The boosted Coulomb dressings in equation (\ref{v coulomb dressing})
are also special cases of Dirac's dressing prescription (\ref{dirac dressing}) where the boost velocity plays the role of the four-vector $v^\mu$. 
Their results are consistent with what we have found above, together with one additional  interesting 
observation. They noted that the mass-shell logarithms cancel identically when they fine-tune the velocity of the photon cloud to match the momentum,   
 $v^\mu=\frac{1}{m} p^\mu$. This velocity tuning is directly analogous to the same in a Faddeev-Kulish asymptotic state (\ref{FK}).  It is also a feature of our result for Wilson-line-dressed particles where the exponent $\gamma(\zeta)$ in equation (\ref{infraparticle exponent 1}) vanishes when $\zeta=0$, again the limit where $v^\mu=\frac{1}{m}p^\mu$.    
 
 It is tempting to identify the dressed state with velocity tuned to the momentum as one which has a single isolated charged particle in its spectrum.  However, this is incompatible with rigorous theorems, particularly Buchholz theorem \cite{Buch}, which imply that such a state cannot be gauge invariant and carry a unitary representation of the Lorentz group at the same time.  Also, we have seen that $\gamma(\zeta)$ vanishes only on the single ray $p\propto v$, a set of zero measure in the spectrum of the state $\Phi^*_v\ket{0}$, so no discrete mass eigenvalue arises. 
 
 Another  important observation that Bagan, Lavelle and McMullan made is that at least a certain class of dressings are equivalent to particular choices of gauge.   For example, with the Coulomb gauge, where $\vec\nabla\cdot\vec A(x)=0$, the dressed operator in (\ref{v coulomb dressing}) with $\vec v=0$ is identical to the undressed operator.   It was known before their work that, upon computing the electron propagator in the non-covariant Coulomb gauge, the on-shell infrared (IR) divergences canceled when one set $p^\mu = (m, \vec{0})$ \cite{Zumino:1960}.  Bagan, Lavelle and McMullan showed that this is rigorously so for the Coulomb gauge and also for the boosted Coulomb dressings that they studied. 
 
 The reader might wonder, once we have learned that $v^\mu$ is superselected and all Wilson line dressings must have the same $v$, whether the dressing that we have used in (\ref{dress}) is simply equivalent to using undressed operators in a particular gauge where the gauge condition is $v^\mu A_\mu(x)=0$. In the Euclidean computations,  this would be equivalent to an axial gauge.  Going to the axial gauge is subtle in that the gauge transformation that is needed does not generally fall off asymptotically.
A symptom of this fact are the eikonal poles $\sim\frac{1}{{v\cdot k}}$
and $\sim\frac{1}{({v\cdot k})^2}$ in the axial gauge propagator \cite{Lieb,Lieb1},
\begin{equation}
\Delta_{\mu\nu}(k)=\frac{1}{k^2}\left(\delta_{\mu\nu}-\frac{v_\mu k_\nu+v_\nu k_\mu}{v\cdot k}+\frac{v^2k_\mu k_\nu}{(v\cdot k)^2}\right)
\end{equation}
We shall find that the Wilson line dressing is indeed very similar to the axial gauge but where the distributional properties of the eikonal pole $\frac{1}{v\cdot k}$ and the eikonal double pole $\frac{1}{(v\cdot k)^2}$ are uniquely specified by the detailed nature of the dressing.

The infraparticle exponent in equation (\ref{infraparticle exponent 1}) should be visible in the spacetime behaviour of states that are created by the dressed operators.   For example, consider the state $\Phi_v^*(0)\ket{0}$ and ask what is the overlap with the state $\Phi_v^*(x)\ket{0}$. This is the amplitude $\mathcal A$ that the particle, after time $x^0$, is  located in the vicinity of $\vec x$. For the moment, we will ignore the subtlety that these states should be suitably smeared so that they are normalizable. 

If it were a free relativistic particle with mass $m$,  and $t=\sqrt{-x^\mu x_\mu}$ is the proper time of the propagation, in the large $mt\gg1$ limit, 
\begin{equation}
\begin{aligned}
\mathcal A_{\rm particle}(t)~\sim~\frac{ e^{-imt}}{t^\frac{3}{2}}
\end{aligned}
\end{equation}
This asymptotic form sets in after $t>\frac{1}{m}$ which is typically a very short time -- the Compton time of the electron is $8.09\times10^{-21}sec$   and the Compton time of the proton is $4.41\times10^{-24}sec$. 

For the infraparticle, on the other hand, the long time limit is governed by the edge of the cut singularity in the propagator and the excitation diffuses as
\begin{equation}\label{infraparicle scaling}
\begin{aligned}
&\mathcal A_{\rm infraparticle}(t)\sim \frac{ e^{-imt}}{t^{\frac{3}{2}+\gamma(\zeta)}}
\\
&\gamma(\zeta)=\frac{e^{2}}{2\pi^{2}}\bigl(\zeta\coth\zeta-1\bigr)
\\&
\cosh\zeta =- \frac{v^\mu x_\mu}{\sqrt{-x^\mu x_\mu}}~,~~t=\sqrt{-x^\mu x_\mu}
\end{aligned}
\end{equation}

The exponent $\gamma(\zeta)$
 is the same function as in equation (\ref{infraparticle exponent 1}). The rapidity 
$\zeta$, defined there through the momentum, is defined here through the direction of propagation; the two coincide at late times, where the amplitude is dominated by the classical trajectory $p_\mu=mx_\mu/t$.
This exponent  is computable in perturbation theory, is one-loop exact (for the same reason) and it depends on the dressing of the field that created the excitation through its dependence on the rapidity $\zeta$.   

This gives us the remarkable conclusion that there is an exactly computable exponent governing the spreading of the position  of a charged particle.  The exponent is not quite universal in the sense that it depends on the dressing which one could regard, at least in a loose sense, as the record of how the charged particle was prepared. This brings up the fascinating question as to whether the infraparticle exponent and its dressing dependence could be measurable by experiment.  We leave this question for future work.

We have already noted that the photon cloud superselection rule is an idealization which becomes significantly less constraining if the photon has even a tiny physical mass, or the spacetime has finite extent. The existence of the infraparticle also relies on the idealizations of an exactly massless photon propagating in open, infinite spacetime.  If the photon had a small physical mass, as we shall argue in section \ref{model}, there are local gauge invariant operators given in equation (\ref{gifs}) made with the St\"uckelberg field, which create charged states.  The soft dressing that we are discussing here would not be needed.   On the other hand, in a scenario where the photon is exactly massless and the St\"uckelberg dressed local operators are not available,  one might give the photon a mass anyway to emulate finite time or finite volume effects.  We would still require the gauge invariant dressing of charged operators that we have been discussing.  In that case, absolute cloud orthogonality is lost.  As well the infraparticle is lost.  It reverts to the more conventional scenario where the 2 point function has a pole singularity at $p^2=-m^2$, the physical mass of the charged particle, and then a cut singularity beginning at the particle plus photon threshold $p^2=-(m+m_\gamma)^2$.

On the other hand, as we will show in section \ref{window}, even if the photon has a small mass $m_\gamma$, the power law scaling of the amplitude in equation (\ref{infraparicle scaling}) with the same exponent $\gamma(\zeta)$ persists for a large range of times.\footnote{A way to see this is to simply note that, without remarkably good energy resolution, it would be impossible to distinguish the infraparticle behaviour in the vicinity of $-p^2\sim m^2$ from the conventional pole at $-p^2=m^2$ and the cut beginning at the threshold $(m+m_\gamma)^2$.  This energy resolution would require times of order $1/m_\gamma$ which is already a macroscopic time interval if the photon Compton wave-length is of astronomical magnitude.}  It should set in when the asymptotic expansion of the 2 point function becomes valid, when $t>>1/m$.  This is a short time, the Compton time of the massive charged particle. We will show that it is then valid up to time $t\lessapprox 1/m_\gamma$, the Compton time of the photon which by current experimental bounds is at least of the order of hours.  We find that, even in the non-idealized system, this scaling formula is still valid over this remarkable dynamical range.  We call this range the infrared window $\frac{1}{m}\lesssim t\lesssim \frac{1}{m_\gamma}$.  

What is more, as is depicted in figure \ref{fig:crossover}, the crossover between the $t\lessapprox 1/m_\gamma$ anomalous scaling and the late time $t\gtrapprox 1/m_\gamma$ classical scaling is rather narrow.   In any realistic scenario, the main limitation to seeing this scaling law would not be the width of the infrared window, rather  other sources of decoherence, for example,  from the environment or from photoemission, which we have not analyzed here.  

  The remainder of this paper is organized as follows.
In section \ref{model} we will prepare  the infrared cutoff of quantum electrodynamics by giving a careful gauge invariant formulation of scalar electrodynamics with a photon mass. In section \ref{self energy} we give the details of the one loop order computation of the undressed 2 point functions of the scalar fields. In section \ref{wilson lines} we complete the computation of the 2 point function of dressed operators by including correlations between the particle and the Wilson line dressings and the self-correlations of the dressings. In section \ref{window} we discuss the infrared window where, even if the photon has a small mass, the proper time propagator has the infraparticle scaling.  In section \ref{conclusion} we summarize some further remarks.

\section{Infrared Cutoff Scalar QED}
\label{model}
We will study the properties of correlation functions of dressed operators in scalar quantum electrodynamics and using perturbation theory in  the electric charge $e$.  Scalar QED  is chosen for its relative technical simplicity.   Our results should hold in spinor electrodynamics or more complicated theories with multiple U(1) gauge groups or multiple species of charged fields as well.
As an ultraviolet regulator, we will use dimensional regularization throughout by taking the spacetime dimension to be $D=4-2\epsilon$ and eventually evaluating results in the regime where $\epsilon\sim0$.

\subsection{BRST invariant QED with a Photon mass}

As an infrared regulator, we will introduce a fundamental photon mass which we denote by $m_\gamma$. We will make the photon massive while  retaining the gauge symmetry by using the St\"uckelberg mechanism.
A review of the mechanism for QED and other gauge field theories can be found in reference \cite{Ruegg:2003ps}.  Throughout we shall assume that
$0\leq m_\gamma<2m$ where $m$ is the mass of the charged scalar field, so that the photon is a stable particle.
The Euclidean action of scalar QED with a St\"uckelberg mass is given by
\begin{equation}\label{S}
\begin{aligned}
{\bf S}=\int d^{D}x~ & \biggl\{|(\partial_\mu-ieA_\mu(x))\phi(x)|^{2}+m^{2}|\phi(x)|^{2}
+\frac{\lambda}{2}|\phi(x)|^{4}
\\&
+\frac{1}{4}F_{\mu\nu}^{2}(x)
+\frac{m_\gamma^{2}}{2}\Big(A_\nu(x)-\frac{1}{m_\gamma}\partial_\nu\varphi(x)\Big)^{2}
\biggr\}
\end{aligned}
\end{equation}
The action is   invariant  under the gauge transformations
\begin{equation}\label{gt}
\begin{aligned}
&\phi(x)\to e^{ie\eta(x)}\phi(x)
~,~~
\phi^*(x)\to e^{-ie\eta(x)}\phi^*(x)
\\&
A_\mu(x)\to A_\mu(x)+\partial_\mu\eta(x)
~,~~\varphi(x)\to \varphi(x)+m_\gamma\eta(x)
\end{aligned}
\end{equation}
We also require a linearly realized global U(1) symmetry that transforms the
complex scalar fields
\begin{equation}\label{global u(1)}
\begin{aligned}
&\phi(x)\to e^{ie\theta}\phi(x)
~,~~
\phi^*(x)\to e^{-ie\theta}\phi^*(x)
\end{aligned}
\end{equation}
We have not included the transformation $\varphi\to \varphi+m_\gamma \theta$ in the global U(1) transformations (\ref{global u(1)}).  Such a transformation could not be implemented by a proper unitary operation when the spacetime dimension is greater than two.

The marginal operator with the quartic coupling $\lambda|\phi|^4$ is required for
renormalizability.  Its natural order is $\lambda \sim e^{4}$ and we shall assume that it is
of this order, so that it will play no role in our one loop, order $e^2$ computations.

Two further operators, $$ \Big(A_\nu-\frac{1}{m_\gamma}\partial_\nu\varphi\Big)^2|\phi|^2~~,~~\left[\Big(A_\nu-\frac{1}{m_\gamma}\partial_\nu\varphi\Big)^2\right]^2 $$
are compatible with the symmetries and are marginal.  We do not include them as, if they are put to zero at the outset, they are not required for renormalizability.  This is due to the fact that, with them set to zero and after gauge fixing, we will  find that, the St\"uckelberg field is a decoupled free field. As a consequence, there are no interactions which could generate such terms.

Other non-polynomial local operators containing $\varphi(x)$ such as
\begin{equation}\label{gifs}
\phi(x)~e^{-ie\varphi(x)/m_\gamma}
~~,~~~
  \phi^*(x)~e^{ie\varphi(x)/m_\gamma}
  \end{equation}
 are gauge invariant.   (We will make extensive use of these in a different context shortly.)  We have specifically chosen the action of the global U(1) symmetry (\ref{global u(1)}) so that these operators are U(1) charged.  This eliminates some monomials made from these gauge invariant, but not phase invariant, operators from the action.

In the quantum field theory that is described by the action (\ref{S}), the only physically meaningful observables are  gauge invariant operators.  With this in mind, we can remove the gauge redundancy by fixing a gauge.   A Lorentz covariant gauge fixing is implemented by adding the gauge fixing action
\begin{equation}\label{sgf}
{\bf S}_{g.f.}=\int d^{D}x~\left\{\frac{1}{2\xi}B^{2}(x)
+iB(x)\Big(\partial_\mu A_\mu(x)-\frac{m_\gamma}{\xi}\varphi(x)\Big)
+\partial_\mu b(x) \partial_\mu c(x)+\frac{m_\gamma^{2}}{\xi}b(x)c(x)\right\}
\end{equation}
where $B(x)$ is the Nakanishi-Lautrup field and $b(x)$ and $c(x)$ are Faddeev-Popov ghost fields.
This gauge fixing action is exact under a nilpotent Grassmann odd derivation, the BRST transformation, which acts on the fields as
\begin{equation}\label{BRST}
\begin{aligned}
&s:\phi(x)=iec(x)\phi(x)
~,~s:\phi^{*}(x)=-iec(x)\phi^{*}(x)
~,~s:A_\mu(x)=\partial_\mu c(x)
~,~s:\varphi(x)=m_\gamma c(x)
\\ &
s:c(x)=0
~,~
s:b(x)=iB(x)
~,~s:B(x)=0
\end{aligned}
\end{equation}
Indeed, we can see that the gauge fixing action is BRST exact
\begin{equation}\begin{aligned}
&{\bf S}_{g.f.}~=~s:~\Psi
\end{aligned}\end{equation}
where the gauge-fixing fermion is
\begin{equation}\begin{aligned}
\Psi=\int d^{D}x~b(x)\left[-i\frac{B(x)}{2\xi}+\partial_\mu A_\mu(x)
-\frac{m_\gamma}{\xi}\varphi(x)\right]
\end{aligned}\end{equation}
Correlation functions of BRST invariant
operators are therefore independent of the gauge fixing, particularly the parameter $\xi$.

 After integrating out the Nakanishi--Lautrup field
  and the decoupled ghosts, perturbation theory is governed  by the gauge fixed action
\begin{equation}\label{Sprime}
\begin{aligned}
{\bf S}'=\int d^{D}x&\biggl\{|(\partial_\mu-ieA_\mu (x) )\phi(x) |^{2}+m^{2}|\phi(x) |^{2}
+\frac{\lambda}{2}|\phi(x) |^{4}+\frac{1}{4}F_{\mu\nu}^2(x)+\frac{m_\gamma^2}{2}A_\mu^2(x)
\\&+\frac{\xi}{2}(\partial_\mu A_\mu(x) )^2
 + \frac{1}{2}(\partial_\mu\varphi(x))^2 +\frac{m_\gamma^2}{2\xi}\varphi^2(x)
 \biggr\}
\end{aligned}
\end{equation}
The St\"uckelberg scalar field has also decoupled and it could also be integrated out.  However, shortly, we shall
find a use for retaining it.  We see that when $\xi\to0$ (and with $\varphi(x)$ decoupled), this is the standard action which describes a massive Proca field, in this case coupled to the massive complex scalar field and possessing a global U(1) symmetry.  We will use the Feynman gauge, $\xi=1$, for most of our computations.

Once we have fixed the gauge, expectation values of gauge (and therefore BRST)-invariant operators $\mathcal O$ are to be computed with the Euclidean  functional integration
\begin{equation}
\left<~\mathcal O~\right>~~\equiv~~ \frac{ \bigintss[d\phi dAd\varphi]e^{-{\bf S}'[A,\phi,\varphi]}~~\mathcal O~}
{\bigintss[d\phi dAd\varphi]e^{-{\bf S}'[A,\phi,\varphi] } }
\end{equation}

\subsection{St\"uckelberg correlations}

We will be interested in the correlation functions of gauge invariant operators which carry the global U(1) charge.
As long as the photon mass $m_\gamma$ is nonzero, the  local composite operators listed in equation (\ref{gifs}) are gauge invariant and carry U(1) charges. We could consider a multipoint correlation function of these gauge invariant operators,
\begin{equation}\label{factorization 1}
\begin{aligned}
&\left< ~\phi(x_1)e^{-ie\varphi(x_1)/m_\gamma}~\ldots~\phi(x_n)e^{-ie\varphi(x_n)/m_\gamma}~
\phi^*(y_1)e^{ie\varphi(y_1)/m_\gamma}~\ldots~\phi^*(y_n)e^{ie\varphi(y_n)/m_\gamma}~
\right>
\\&=
\left< \phi(x_1)\ldots\phi(x_n)
\phi^*(y_1)\ldots\phi^*(y_n)
\right>
\left< e^{-ie\varphi(x_1)/m_\gamma}\ldots e^{-ie\varphi(x_n)/m_\gamma}
e^{ie\varphi(y_1)/m_\gamma}\ldots e^{ie\varphi(y_n)/m_\gamma}
\right>
\\&
=
\left< \phi(x_1)\ldots\phi(x_n)
\phi^*(y_1)\ldots\phi^*(y_n)
\right>
e^{
-\frac{e^2}{2m_\gamma^2}\sum\limits_{i,j=1}^n
\left[g(x_i,x_j)+g(y_i,y_j)-2g(x_i,y_j)\right]}
 \\&
 ~~~g(x,y)~\equiv~(x|\frac{1}{-\partial^2+m_\gamma^2/\xi}|y)
 \end{aligned}
\end{equation}
where the factorization is due to the fact that, in ${\bf S}'$ in (\ref{Sprime}),  $\varphi$ is a decoupled free field.
We have used this to compute the second factor.
This factorization of the correlation function, arising from triviality of the St\"uckelberg scalar field has interesting consequences.  For example,
from the last equality in (\ref{factorization 1}) we see that the general multipoint function of the undressed complex scalar fields $\left< \phi(x_1)\ldots\phi(x_n)\phi^*(y_1)\ldots\phi^*(y_n)\right>$ in gauge-fixed QED with a massive photon is equal to a gauge invariant correlation function of gauge invariant operators times a classical gauge variant function whose exact form and gauge parameter dependence we know.
\begin{equation}\label{lkf}
\begin{aligned}
&\left< \phi(x_1)\ldots\phi(x_n)
\phi^*(y_1)\ldots\phi^*(y_n)
\right>
\\&=\left< \phi(x_1)e^{-ie\varphi(x_1)/m_\gamma}\ldots\phi(x_n)e^{-ie\varphi(x_n)/m_\gamma}
\phi^*(y_1)e^{ie\varphi(y_1)/m_\gamma}\ldots\phi^*(y_n)e^{ie\varphi(y_n)/m_\gamma}
\right>
\times
\\&
\times
e^{
\frac{e^2}{2m_\gamma^2}\sum\limits_{i,j=1}^n
\left[g(x_i,x_j)+g(y_i,y_j)-2g(x_i,y_j)\right]}~,~~g(x,y)=\int\frac{d^Dk}{(2\pi)^D}\frac{e^{ik(x-y)} }{k^2+m_\gamma^2/\xi}
\end{aligned}\end{equation}
This gives us what is in essence an elementary derivation of the Landau-Khalatnikov-Fradkin transformation \cite{lan1,frad1}  for scalar QED with a finite infrared cutoff. The Landau-Khalatnikov-Fradkin transformation relates gauge variant correlation functions of undressed fields in different relativistic gauges, a $\xi$-gauge and a $\xi'$-gauge, in this instance,
  \begin{equation}
  \begin{aligned}
& \left< \phi(x_1)\ldots\phi(x_n)
\phi^*(y_1)\ldots\phi^*(y_n)
\right>_\xi=
\\&
\left< \phi(x_1)\ldots\phi(x_n)
\phi^*(y_1)\ldots\phi^*(y_n)
\right>_{ \xi '}
e^{\frac{e^2}{2}\left(\frac{1}{ \xi '}-\frac{1}{\xi}\right) \sum\limits_{i,j=1}^n
\left[\Delta(x_i,x_j)+\Delta(y_i,y_j)-2\Delta(x_i,y_j)\right]}
\\ &
\Delta(x)=\mu^{4-D}\int \frac{d^Dk}{(2\pi)^D}\frac{e^{ikx}}{[k^2+\frac{1}{\xi}m_\gamma^2][k^2+\frac{1}{ \xi '}m_\gamma^2]}
\end{aligned}
\end{equation}

The coincident-point Green functions, $(x_i|\tfrac{1}{-\partial^2+m_\gamma^2/\xi}|x_i)$, in the exponent in (\ref{lkf}) are ultraviolet divergent.  Assuming that the 2n point function $\left< \phi(x_1)\ldots\phi(x_n)\phi^*(y_1)\ldots\phi^*(y_n)\right>$ has been rendered finite by the usual renormalization machinery of a renormalizable quantum field theory, these extra ultraviolet divergences can be  absorbed by the multiplicative renormalization of the composite operators (\ref{gifs}). This yields a proof of renormalizability of the non-polynomial local composites
$\phi(x)~e^{-ie\varphi(x)/m_\gamma}$ and $\phi^*(x)~e^{ie\varphi(x)/m_\gamma}$.

\subsection{If the photon has a mass}

Quantum electrodynamics with a massive photon  could be interesting for the description of electrodynamics in nature, as there we do not know that the photon is precisely massless, there are only experimental upper bounds on its mass.  Those bounds are very stringent -- the Compton wavelength of the photon is at least of solar system distance scales -- but they do not put the mass at zero. 

Of course, if the physical photon and quantum electrodynamics actually has a photon with a small mass, the local St\"uckelberg operators  (\ref{gifs})  can be used to create charged states.  Our study of dressed operators would not be necessary.  On the other hand, we might prefer the aesthetic of setting the photon mass precisely to zero in the QED Lagrangian.
In that case, our construction of gauge invariant local operators in (\ref{gifs}) fails.  Quantum electrodynamics with a massless photon does not have a local gauge invariant U(1) charged operator and we are back to non-local dressing. 

There could be the scenario where the photon in the Lagrangian density of QED is massless and we are required to dress local charged operators, but we use QED with a massive photon anyway, to emulate some other cutoff, such as finiteness of the volume of space, or finiteness of the proper time since the big bang.  

In that case, the charged states must be created by a dressed operator but we no longer remove the infrared cutoff.  We simply put $m_\gamma$ to its small physical value. 
The small photon  mass enters expressions such as $\frac{e^2}{2\pi}\ln\frac{m_\gamma}{m}$ where $\frac{e^2}{2\pi}$ is itself a small number. For almost any conceivable values of $m_\gamma/m$, this contribution, rather than being logarithmically divergent, is itself  rather small.   

To see this, If we use the current upper bound for the photon mass $m_\gamma$ and  the electron mass for $m=m_{\rm el}$,  the suppression factor in the photon cloud orthogonality formula (\ref{cusp 1}) is
\begin{equation}\label{sudakov}
\left(\frac{m^2_\gamma}{m_{\rm el}^2}\right)^{\frac{e^2}{8\pi^2}(\chi\coth\chi-1)} ~~\approx~~(.88)^{(\chi\coth\chi-1)}
\end{equation}
If we let the photon Compton wavelength be the Hubble radius, or the Compton time be the proper
time since the big bang, we would have
\begin{equation}\label{sudakov 1}
\left(\frac{m^2_\gamma}{m_{\rm el}^2}\right)^{\frac{e^2}{8\pi^2}(\chi\coth\chi-1)} ~~\approx~~(.814)^{(\chi\coth\chi-1)}
\end{equation}
These factors, rather than being dramatic,  would  be significant at all only for $\chi$ in the ultra-relativistic regime, $\chi\gg1$ where $(\chi\coth\chi-1)\approx\chi$. The attenuation would be at a 20 percent level when $\chi\sim 3$, which is a relative speed of $0.995 $c. 

In this scenario, photon clouds still exist but cloud orthogonality goes away and is replaced by a milder damping.  Similarly, the infraparticle also goes away.  The intermediate states in the 2 point function revert to a more conventional  isolated massive charged particle state, which gives the 2 point function its pole, and the particle plus photon states which give it its cut singularity.   However, we argue in the introduction, and we show in section \ref{window} that there is a significant window of proper time where, if the photon is very light, the infraparticle scaling persists.  This spans times longer than the Compton time of the charged particle up to the Compton time of the photon.

\section{The undressed self-energy and the residue obstruction}
\label{self energy}

\subsection{One loop self energy of the complex scalar field}

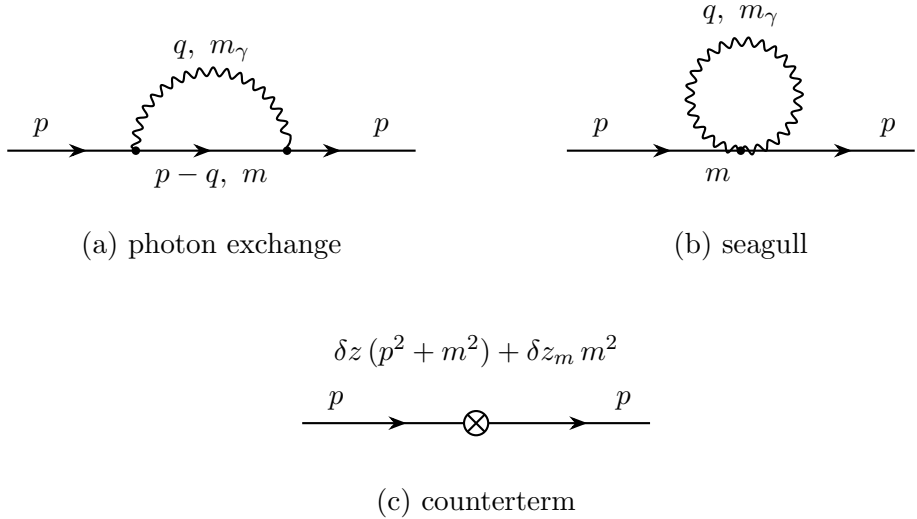
\begin{figure}[h]
\centering
\begin{tikzpicture}[
  >={Stealth[length=2.4mm]},
  photon/.style={thick,decorate,
     decoration={snake,amplitude=1.6pt,segment length=5pt}},
  charge/.style={thick,postaction=decorate}
]
\begin{scope}
\coordinate (A) at (-1,0);
\coordinate (B) at ( 1,0);
\draw[charge,decoration={markings,mark=at position 0.62 with {\arrow{>}}}]
   (-2.7,0) -- (A);
\draw[charge,decoration={markings,mark=at position 0.5 with {\arrow{>}}}]
   (A) -- (B);
\draw[charge,decoration={markings,mark=at position 0.42 with {\arrow{>}}}]
   (B) -- (2.7,0);
\draw[photon] (A) arc (180:0:1);
\fill (A) circle (1.6pt);
\fill (B) circle (1.6pt);
\node at (-2.25,0.30) {$p$};
\node at ( 2.25,0.30) {$p$};
\node at (0,-0.34) {$p-q,\ m$};
\node at (0, 1.32) {$q,\ m_\gamma$};
\node at (0,-1.25) {(a) photon exchange};
\end{scope}
\begin{scope}[xshift=7cm]
\coordinate (V) at (0,0);
\draw[charge,decoration={markings,
      mark=at position 0.30 with {\arrow{>}},
      mark=at position 0.82 with {\arrow{>}}}]
   (-2.3,0) -- (2.3,0);
\draw[photon] (V) arc (-90:270:0.72);
\fill (V) circle (1.6pt);
\node at (-1.85,0.30) {$p$};
\node at ( 1.95,0.30) {$p$};
\node at (-0.30,-0.32) {$m$};
\node at (0,1.78) {$q,\ m_\gamma$};
\node at (0,-1.25) {(b) seagull};
\end{scope}
\begin{scope}[xshift=3.5cm,yshift=-3.6cm]
\coordinate (C) at (0,0);
\draw[charge,decoration={markings,
      mark=at position 0.30 with {\arrow{>}},
      mark=at position 0.82 with {\arrow{>}}}]
   (-2.3,0) -- (2.3,0);
\draw[thick,fill=white] (C) circle (0.16);
\draw[thick] (-0.113,-0.113) -- (0.113,0.113);
\draw[thick] (-0.113,0.113) -- (0.113,-0.113);
\node at (-1.85,0.30) {$p$};
\node at ( 1.95,0.30) {$p$};
\node at (0,0.95) {$\delta z\,(p^{2}+m^{2})+\delta z_m\,m^{2}$};
\node at (0,-1.05) {(c) counterterm};
\end{scope}
\end{tikzpicture}
\caption{The one-loop diagrams contributing to the charged-scalar
self-energy $\Pi(p^2)$:  (a) the photon-exchange diagram, (b) the seagull diagram (c) the counterterm insertion.  Solid oriented lines are the complex scalar, wavy lines are the photon. }
\label{fig:diagrams}
\end{figure}

The two point function of the undressed field is
\begin{equation}
\braket{ \phi(x)\phi^{*}(0) }
=\int\frac{d^{D}p}{(2\pi)^{D}}\frac{ e^{ipx}}{p^{2}+m^{2}+\Pi(p^{2})}
\end{equation}
where the self-energy $\Pi(p^2)$ is (minus) the sum of amputated one-particle-irreducible graphs,
including the counterterm contribution
$$
\delta z\,(p^{2}+m^{2})+\delta z_m
m^{2}
$$
In the one-loop approximation the self-energy of the complex scalar field
receives a contribution from the photon-exchange diagram and from the seagull
(``tadpole'') diagram, shown in figure~\ref{fig:diagrams}.
The leading one loop contribution to the scalar self-energy corresponding to the diagrams in figure \ref{fig:diagrams} is given by the Feynman integral
\begin{equation}
\Pi(p^2) = e^2\mu^{4-D}\!\int\!\frac{d^Dq}{(2\pi)^D}
\left[\frac{D}{q^2+m_\gamma^2}
-\frac{(2p-q)^2}{[(p-q)^2+m^2]\,[q^2+m_\gamma^2]}\right]
+\delta z\,(p^{2}+m^{2})+\delta z_m\,m^{2} ,
\label{eq:Pi-start}
\end{equation}
where  $\mu$ is the 't~Hooft-Veltman mass that keeps $e$ dimensionless away from $D=4$.
The integration is elementary. The counterterms must be determined so that they remove the ultraviolet singularities. In the $\overline{\bf MS}$ subtraction scheme, the counterterms are determined as
\begin{equation}\label{MSbarCT}
\begin{aligned}
&
\delta z_{\overline{\rm MS}}=\frac{e^{2}}{8\pi^{2}\bar\epsilon}
\\&
\delta z_{m,\overline{\rm MS}}=-\frac{3e^{2}}{16\pi^{2}}\Big(1-\frac{m_\gamma^{2}}{m^{2}}\Big)
\frac{1}{\bar\epsilon}\
\\&\frac{1}{\bar\epsilon}~\equiv~\frac{1}{\epsilon}-\gamma_{E}+\ln4\pi
\end{aligned}
\end{equation}
with $\gamma_{E}\approx 0.57721 ...$ the Euler-Mascheroni constant.
The minimally subtracted renormalized self-energy is given by the expression
\begin{equation}\label{PiMSbar}
\begin{aligned}
\Pi_{\overline{\rm MS}}(p^{2})=&\frac{e^{2}}{16\pi^{2}}\biggl\{
\big(2p^{2}-2m^{2}+m_\gamma^{2}\big)\frac{\lambda_\gamma}{p^{2}}
\ln\frac{ p^{2}+m^{2}+m_\gamma^{2}+\lambda_\gamma}{2mm_\gamma}
+\Big(p^{2}-m^{2}+\tfrac32 m_\gamma^{2}\Big)\ln\frac{m^{2}}{\mu^{2}}\\
&+\Big(p^{2}+\tfrac32 m_\gamma^{2}\Big)\ln\frac{m_\gamma^{2}}{\mu^{2}}
+\frac{(m^{2}-m_\gamma^{2})(2m^{2}-m_\gamma^{2})}{2p^{2}}
\ln\frac{m^{2}}{m_\gamma^{2}}-4p^{2}+3m^{2}-2m_\gamma^{2}\biggr\}
\end{aligned}
\end{equation}
where
\begin{equation}\label{kallen root}
\lambda_\gamma~\equiv~\sqrt{(p^{2}+(m+m_\gamma)^{2})(p^{2}+(m-m_\gamma)^{2})}
\end{equation}
is the K\"all\'en root.
The expression in equation (\ref{PiMSbar}) remains finite as the infrared cutoff is removed, $m_\gamma\to 0$. However, 
that limit does not commute with the mass shell limit $p^2\to-m^2$. If we put $p^2\to -m^2$ first, 
its derivative at the mass shell, the residue, becomes logarithmically divergent
as $m_\gamma\to 0$.  

In the two limits $\ln(p^2+m^2)$ at $m_\gamma\to 0$ is being traded for $\ln(m_\gamma^2)$ at $p^2\to-m^2$.  This is the residue obstruction.  It does not allow renormalization of the undressed 2 point function so that it has
a pole.
 If we nevertheless, while keeping $m_\gamma$ small but nonzero, implement  the on-shell subtraction scheme, the counterterms are determined by the conditions
$$
\Pi_{\rm pole}(-m^{2})=\Pi_{\rm pole}'(-m^{2})=0
$$
The on-shell (pole) subtraction counterterms are equal to the minimal subtraction counterterms  shifted by ultraviolet finite functions of the parameters,
\begin{equation}\label{ABconsts}
\begin{aligned}
&\delta z_{\rm pole}\to\delta z_{\overline{\rm MS}}-\frac{e^{2}}{16\pi^{2}}B
\\& \delta z_{m,{\rm pole}}\to\delta z_{m,\overline{\rm MS}}-\frac{e^{2}}{16\pi^{2}}\frac{A}{m^{2}}
\\& A=-\frac{m_\gamma(4m^{2}-m_\gamma^{2})^{3/2}}{m^{2}}
\arctan\frac{\sqrt{4m^{2}-m_\gamma^{2}}}{m_\gamma}
-3(m^{2}-m_\gamma^{2})\ln\frac{m^{2}}{\mu^{2}}
-\frac{m_\gamma^{4}}{2m^{2}}\ln\frac{m^{2}}{m_\gamma^{2}}
+7m^{2}-2m_\gamma^{2}
\\& B=-\frac{m_\gamma(m^{2}-m_\gamma^{2})\sqrt{4m^{2}-m_\gamma^{2}}}{m^{4}}
\arctan\frac{\sqrt{4m^{2}-m_\gamma^{2}}}{m_\gamma}
+\frac{m_\gamma^{2}(3m^{2}-m_\gamma^{2})}{2m^{4}}\ln\frac{m^{2}}{m_\gamma^{2}}
+2\ln\frac{m_\gamma^{2}}{\mu^{2}}-\frac{m_\gamma^{2}}{m^{2}}
\end{aligned}
\end{equation}
and the on-shell-subtracted self-energy is
\begin{equation}\label{Pipole}
\begin{aligned}
&\Pi_{\rm pole}(p^{2})=\frac{e^{2}}{16\pi^{2}}\biggl\{
\big(2p^{2}-2m^{2}+m_\gamma^{2}\big)\frac{\lambda_\gamma}{2p^{2}}
\ln\frac{\big(p^{2}+m^{2}+m_\gamma^{2}
+\lambda_\gamma\big)^{2}}{4m^{2}m_\gamma^{2}}
\\&
+\frac{(m^{2}-m_\gamma^{2})(2m^{2}-m_\gamma^{2})(p^{2}+m^{2})^{2}}{2p^{2}m^{4}}
\ln\frac{m^{2}}{m_\gamma^{2}}-\frac{(p^{2}+m^{2})(4m^{2}-m_\gamma^{2})}{m^{2}}
\\
&+\frac{m_\gamma\sqrt{4m^{2}-m_\gamma^{2}}}{m^{4}}
\Big[m^{2}(4m^{2}-m_\gamma^{2})+(p^{2}+m^{2})(m^{2}-m_\gamma^{2})\Big]
\arctan\frac{\sqrt{4m^{2}-m_\gamma^{2}}}{m_\gamma}
\biggr\}
\end{aligned}
\end{equation}
While $\Pi_{\overline{\rm MS}}$ has a finite limit as the infrared regulator is removed, $m_\gamma\to0$,
\begin{equation}
\lim_{m_\gamma\to0}\Pi_{\overline{\rm MS}}(p^{2})=
\frac{e^{2}(p^{4}-m^{4})}{8\pi^{2}p^{2}}\ln\frac{p^{2}+m^{2}}{m^{2}}
-\frac{e^{2}}{16\pi^{2}}\Big[(2p^{2}-m^{2})\ln\frac{\mu^{2}}{m^{2}}
+4p^{2}-3m^{2}\Big]
\end{equation}
the on-shell-subtracted self-energy does not:
\begin{equation}\label{PipoleIR}
\begin{aligned}
\Pi_{\rm pole}(p^{2})&~~\overset{m_\gamma\to0}{\longrightarrow}~~
\frac{e^{2}(p^{4}-m^{4})}{8\pi^{2}p^{2}}\ln\frac{p^{2}+m^{2}}{m^{2}}
-\frac{e^{2}(p^{2}+m^{2})}{8\pi^{2}}\Big[\ln\frac{m_\gamma^{2}}{m^{2}}+2\Big]
+\mathcal O(m_\gamma)
\end{aligned}
\end{equation}

 The appearance of the  infrared log in $\Pi_{\rm pole}(p^{2})$ is the residue obstruction. 
 For electrodynamics with a massless photon,  the mass shell subtraction scheme results in an infrared
 divergent irreducible 2 point function.
 
In the next section, we will be interested in a hybrid of the two renormalization schemes where we adjust the mass renormalization so that
the pole of the 2 point function occurs at $p^2=-m^2$ -- this is essentially adjusting it so that $\Pi(-m^2)=0$ but we leave the wavefunction renormalization constant to minimally subtract the ultraviolet divergence, that is, we leave it at the $\overline{\rm MS}$ value.  Then we put the infrared cutoff to zero.  What we find is the hybrid subtracted self-energy -- the $m_\gamma\to0$ limit of which is 
\begin{equation}\label{hybrid self energy}
\lim_{m_\gamma\to0}\Pi_{{h}}(p^{2})=[p^2+m^2]
\frac{e^{2}}{8\pi^{2}}\biggl\{
(1-\frac{m^2}{p^{2}})\ln\frac{p^{2}+m^{2}}{m^{2}}
-\Big[\ln\frac{\mu^{2}}{m^{2}}
+2\Big]\biggr\}
\end{equation}
This expression contains the other face of the residue obstruction, the appearance
of a term $\sim \ln[p^{2}+m^{2}]$ which indicates the transmutation of pole to cut singularity.

\subsection{The St\"uckelberg dressed 2 point function}

If the photon had even a tiny physical mass, the St\"uckelberg dressed fields in equation (\ref{gifs}) are local and gauge invariant.  We would not need a nonlocal dressing of charged operators.
As a comparison with the Wilson-line-dressed operators, it is interesting to examine the 2 point function of the St\"uckelberg dressed fields, up to order $e^2$. 
 One point of this exercise is to rule out a
miraculous cancellation of infrared divergences in the limit $m_\gamma\to0$.  Indeed, we shall see
that it does not happen.  In fact, there are power law singularities in that limit, confirming what we might expect, 
that these operators cease to exist in the truly massless photon limit.

As we have already discussed in section \ref{model}, since the St\"uckelberg field is free and decoupled in the gauge-fixed
action, the 2 point function factorizes exactly, to all orders in $e$,  
\begin{equation}\label{St}
\braket{\phi(x)e^{-ie\varphi(x)/m_\gamma}\phi^{*}(0)e^{ie\varphi(0)/m_\gamma}}
=\braket{\phi(x)\phi^{*}(0)} \exp\Big[\frac{e^{2}}{m_\gamma^{2}}\Big(g(x,0)-g(0,0)\Big)\Big]
\end{equation}
We expand to  order $e^2$ as
\begin{equation}\label{gi proca}
\begin{aligned}
&\braket{\phi(x)e^{-ie\varphi(x)/m_\gamma}~ \phi^{*}(0)e^{ie\varphi(0)/m_\gamma}}=
\int \frac{d^Dp}{(2\pi)^D}e^{ipx}D_{\rm S}(p)
\\ &
D_{\rm S}(p) = \frac{1}{p^2+m^2} \biggl[ 1-\frac{\Pi(p^2)}{p^2+m^2}
+ \frac{e^{2}\mu^{4-D} }{m_\gamma^{2} } \int \frac{d^Dk}{(2\pi)^D}  \left[\frac{p^2+m^2}{(p-k)^2+m^2}-1\right]
\frac{1}{k^2+m_\gamma^2}\biggr]+\ldots
\end{aligned}\end{equation}
The one loop integrations are readily done to get
\begin{equation}
\begin{aligned}
&-\frac{e^{2}\mu^{4-D} }{m_\gamma^{2} } \int \frac{d^Dk}{(2\pi)^D}
\frac{1}{k^2+m_\gamma^2}
=\frac{e^2}{16\pi^2}\left[\frac{1}{\epsilon}-\gamma_E+\ln 4\pi+1-\ln\frac{m_\gamma^2}{\mu^2}\right]+\mathcal O(\epsilon)
\\ &
 \frac{e^{2}\mu^{4-D} }{m_\gamma^{2} } \int \frac{d^Dk}{(2\pi)^D}  \frac{p^2+m^2}{(p-k)^2+m^2}\frac{1}{k^2+m_\gamma^2}=
 \frac{e^2(p^2+m^2)}{16\pi^2m_\gamma^2 }\biggl[\frac{1}{\epsilon}-\gamma_E+\ln4\pi +2
\\ &
 -\ln\frac{mm_\gamma}{\mu^2}+\frac{m^2-m_\gamma^2}{2p^2}\ln\frac{m^2}{m_\gamma^2} -\frac{\lambda_\gamma}{p^2}\ln\frac{p^2+m^2+m_\gamma^2+\lambda_\gamma}{2mm_\gamma}\biggr]+\mathcal O(\epsilon)
\end{aligned}\end{equation}
where $\lambda_\gamma$ is the K\"all\'en root defined in equation (\ref{kallen root}). 

Assuming that the undressed self-energy $\Pi(p^2)$ has been made ultraviolet finite, by minimal subtraction for example, the remaining ultraviolet singular terms are
then \begin{align}\label{singular stuckelberg}
D_{\rm S}(p) = \frac{e^2}{16\pi^2m_\gamma^2 }\biggl[\frac{1}{\epsilon}-\gamma_E+\ln4\pi \biggr]
~+~
 \frac{1}{p^2+m^2}   \frac{e^2 }{16\pi^2}\biggl[\frac{1}{\epsilon}-\gamma_E+\ln4\pi \biggr]+\ldots
\end{align}
The first term in (\ref{singular stuckelberg})  is a  pure contact term, in spacetime coordinates it is $\sim \delta(x)/m_\gamma^2$,  and it is not noticeable in $D_{\rm S}(x)$
when $x\neq 0$.  
The second term  on the right-hand-side of (\ref{singular stuckelberg})  must be canceled by multiplicative renormalization of the composite operator. This second term is the same singularity, Taylor expanded in $e^2$ as occurs in the factor (in a more general gauge)
\begin{equation}\label{Stnorm 1}
e^{-\frac{e^{2}}{m_\gamma^{2}}g(0)}
=\exp\Big[\frac{e^{2}}{16\pi^{2}\xi}
\Big(\frac{1}{ \epsilon}-\gamma_E+\ln4\pi+1+\ln\frac{\xi\mu^{2}}{m_\gamma^{2}}\Big)\Big]
\end{equation}
which we also already know must be canceled by multiplicative renormalization, once for each occurrence of  the St\"uckelberg dressed operator in a correlation function.

The full $\bar{\rm MS}$ subtracted result, is
\begin{equation}\label{final 22}
\begin{aligned}
D_{{\rm S}\bar{\rm MS}}(p) =
&\frac{1}{p^{2}+m^{2}}
\left[1+\frac{e^{2}}{16\pi^{2}}
\left(1+\ln\frac{\mu^{2}}{ m_\gamma ^{2}}\right)\right]
-\frac{\Pi_{\overline{\rm MS}}(p^{2})}{(p^{2}+m^{2})^{2}}
\\
&+\frac{e^{2}}{16\pi^{2} m_\gamma ^{2}}
\biggl[2+\ln\frac{\mu^{2}}{m m_\gamma }
+\frac{m^{2}- m_\gamma ^{2}}{2p^{2}}\ln\frac{m^{2}}{ m_\gamma ^{2}}
-\frac{\lambda_\gamma}{p^{2}}
\ln\frac{p^{2}+m^{2}+ m_\gamma ^{2}+\lambda_\gamma}{2m m_\gamma }\biggr]
\\
&+\mathcal O(e^{4})
\end{aligned}
\end{equation}

We can study the limit of the result in  (\ref{final 22})  as the photon mass is put to zero, $m_\gamma\to 0$. 
In the bracket in the second line of (\ref{final 22}) the infrared logarithms cancel among the
three $ m_\gamma $-dependent terms (the coefficient of $\ln m_\gamma $ is
$-1-m^{2}/p^{2}+(p^{2}+m^{2})/p^{2}=0$), leaving a logarithm-free but power law divergent limit,
\begin{equation}
D_{{\rm S}\bar{\rm MS}}(p) ~=~
\frac{e^{2}}{16\pi^{2} m_\gamma ^{2}}
\left[2+\ln\frac{\mu^{2}}{m^{2}}
-\frac{p^{2}+m^{2}}{p^{2}}
\ln\frac{p^{2}+m^{2}}{m^{2}}\right]
+\mathcal O\!\left(\frac{1}{ m_\gamma }\right).
\label{powerdivergence}
\end{equation}
The dressing term therefore diverges as a power, $1/ m_\gamma ^{2}$, at
generic momentum.  This divergence is
nonlocal.  The contribution in (\ref{powerdivergence}) depends on
$p$ through $\ln(p^{2}+m^{2})$, so no contact counterterm, indeed no
local counterterm at all of any dimension, can remove it, and no other term
in the correlator can cancel it.  

The St\"uckelberg-dressed operator
ceases to exist in the $ m_\gamma \to0$ limit, confirming that the local,
isotropic St\"uckelberg cloud is not among the admissible $ m_\gamma \to0$
dressings.

 \section{Including the Wilson lines}
 \label{wilson lines}
 
 In section (\ref{self energy}),  we computed the order $e^2$ correction to the 2 point function of the undressed
 charged scalar field in the form of its irreducible part. Here, we will be interested in the hybrid subtracted function, $\Pi_h(p^2)$ at order $e^2$ which was computed in equation (\ref{hybrid self energy}) of section \ref{self energy}.
 In  the hybrid subtraction scheme, the position of the pole in the 2 point function is indeed at $p^2=-m^2$  (that is, $\Pi_h(p^2=-m^2)=0$) but the residue is minimally subtracted to render it ultraviolet finite.  As we noted in section \ref{self energy}, the result has a finite limit as $m_\gamma$ is put to zero, which is the limit quoted in equation (\ref{hybrid self energy}). The contribution of this irreducible undressed part to the full 2 point function  of the dressed operators will then be $-[p^2+m^2]^{-2}\Pi_h(p^2) $.
 
In this section, we will turn to the 2 point function of the Wilson-line-dressed operators and compute the additional contribution to the 2 point function due to the presence of the Wilson lines.  At order $e^2$, these corrections come in two types, the self-interaction of the Wilson lines and 
the interaction between the particle current and the Wilson lines.  
The perturbative expansion of the Wilson-line-dressed 2 point function is then given by 
\begin{equation}\label{Dv'}
\begin{aligned}
&D_{vv'}(x)~\equiv~
\Braket{ ~e^{+ie\int_0^{\infty} ds v'\cdot A(x+v's)}~\phi(x)~\phi^*(0)~
e^{+ie\int_{-\infty}^{0} ds v\cdot A(vs)}~}
\\
&= \braket{ \phi(x)\phi^*(0) }_0
-\int \frac{d^Dp}{(2\pi)^D} e^{ipx}\frac{ \Pi_h(p)}{[p^2+m^2]^2}
 \\&
 - \braket{ \phi(x)\phi^*(0) }_0  \frac{e^2}{2}
\braket{\left( \int_{-\infty}^0 ds v\cdot A(vs) +\int_0^\infty ds v'\cdot A(x+v's) \right)^2
 }_0
\\&
 + \int d^Dy \int_0^\infty ds  \Braket{ \phi(x)\phi^*(y)}_0 (-ie)\overleftrightarrow{\partial}_\mu
 \Braket{\phi(y)\
 \phi^{*}(0)}_0\Braket{
A_\mu(y)\, (+ie)v'\cdot A(x+v's) }_0
\\ &
+ \int d^Dy \int_{-\infty}^0 ds  \Braket{ \phi(x)\phi^*(y)}_0(-ie)\overleftrightarrow{\partial}_\mu\Braket{\phi(y)\phi^{*}(0)}_0
\Braket{A_\mu(y)\, (+ie)\, v\cdot A(vs) }_0
\\ &
~~~~~~~~~~~~~~~~+~\mathcal O(e^4)
\\ &
 \Braket{ \phi(x)\phi^*(y)}_0=\int \frac{d^Dp}{(2\pi)^D}\frac{e^{ip(x-y)}}{p^2+m^2}
 ~,~
 \Braket{A_\mu(x)A_\nu (y) }_0=\int \frac{d^Dk}{(2\pi)^D}\frac{e^{ik(x-y)}\delta_{\mu\nu}}{k^2+m_\gamma^2}
 \end{aligned}
 \end{equation}
 
  We will use the notation $\delta^{(2)}D_{vv'}(x)$ to denote the order $e^2$, one-loop correction, so that
 \begin{equation}\label{defn of delta2}
 \begin{aligned}
 &D_{vv'}(x)=\int \frac{d^Dp}{(2\pi)^D}\frac{e^{ipx}}{p^2+m^2} +\delta^{(2)}D_{vv'}(x)+\mathcal O(e^4)
 \\
 &\delta^{(2)}D_{vv'}(x) ~\equiv~  \delta_1^{(2)}D_{vv'}(x)+\delta_2^{(2)}D_{vv'}(x)+\delta_3^{(2)}D_{vv'}(x)
\end{aligned}
 \end{equation}
  where $\delta^{(2)}_1D_{vv'}(p)$ is the scalar field self-energy correction appearing in the first line of (\ref{Dv'}), $\delta^{(2)}_2D_{vv'}(p)$ arises from the self-interaction of the Wilson lines appearing in the second line of (\ref{Dv'})  and $\delta^{(2)}_3D_{vv'}(p)$ is due to the interaction between the fields and the Wilson lines, the third and fourth lines of (\ref{Dv'}).
  
 Upon inserting the propagators in (\ref{Dv'}) we find
 \begin{equation}\label{Dv' 2}
 \begin{aligned}
 \delta_1^{(2)}D_{vv'}(p)&=-\frac{ \Pi_{\rm h}(p)}{[p^2+m^2]^2}
\\
 \delta^{(2)}_2D_{vv'}(p)&
 =-\frac{e^2\mu^{4-D}}{2}\int \frac{d^Dk}{(2\pi)^D}\frac{1}{p^2+m^2}
\frac{1}{ k^2+m_\gamma^2}
 \frac{1}{(k\cdot v)^2+\varepsilon^2}+(v\leftrightarrow v')
\\&
+e^2\mu^{4-D}\int \frac{d^Dk}{(2\pi)^D}
\frac{1}{[(p+k)^2+m^2][k^2+m_\gamma^2]}
\frac{v\cdot v'}{(k\cdot v -i\varepsilon)( k\cdot v'-i\varepsilon)}
 \\ \delta^{(2)}_3 D_{vv'}(p)&
 =e^2\frac{2p\cdot v'}{{p}^2+m^2} \int \frac{d^4k}{(2\pi)^4} \frac{1
}{(p+k)^2+m^2}\frac{1}{k^2+m_\gamma^2}\frac{1}{k\cdot v'-i\varepsilon}
+(v'\leftrightarrow v)\\&
+2e^2\mu^{4-D}\frac{1}{{p}^2+m^2} \int \frac{d^Dk}{(2\pi)^D} \frac{1
}{(p+k)^2+m^2}\frac{1}{k^2+m_\gamma^2}\end{aligned}
 \end{equation}

\subsection{Photon cloud superselection: $D_{vv'}(p^2)$}

We know from the results of section \ref{self energy} that $\delta^{(2)}_1D_{vv'}(p)$ in (\ref{Dv' 2}) is  finite as $m_\gamma\to0$.  It is presented in equation (\ref{hybrid self energy}) with that limit already taken.

We can also see from power counting that, at generic values of $p$, the contribution from the interaction of the scalar particle with the Wilson lines, $\delta^{(2)}_3D_{vv'}(p)
$,  remains finite as $m_\gamma\to 0$.

That leaves us with the potential infrared divergences in $\delta^{(2)}_2D_{vv'}(p)
$, which contains the self-interaction of the Wilson lines.   By power counting we expect each of the terms in $\delta^{(2)}_2D_{vv'}(p)$ to be logarithmically divergent as $m_\gamma\to 0$.   Our first task is therefore to find the coefficient of $\ln m_\gamma$ in $\delta^{(2)}_2D_{vv'}(p)$.

Before we isolate the infrared divergence, we need to study the singular product of distributions which appears in the first two terms  in $\delta^{(2)}_2D_{vv'}(p)$ in (\ref{Dv' 2}). That 
product contains   a pinch singularity,
$$
\frac{1}{(k\cdot v)^2+\epsilon^2}=\frac{1}{(k\cdot v -i\varepsilon)}
\frac{1}{( k\cdot v+i\varepsilon)}
$$ 
which we can rewrite as, for example
$$
\frac{1}{(k\cdot v)^2+\epsilon^2}=
\frac{1}{(k\cdot v -i\varepsilon)}
\frac{1}{( k\cdot v+i\varepsilon)}
=\frac{1}{(k\cdot v -i\varepsilon)^2}
+\frac{1}{(k\cdot v -i\varepsilon)}\left[\frac{1}{(k\cdot v +i\varepsilon)}-\frac{1}{(k\cdot v -i\varepsilon)}\right]
$$
$$
=\frac{1}{(k\cdot v -i\varepsilon)^2}
-\frac{2\pi i\delta(k\cdot v)}{(k\cdot v -i\varepsilon)}
=\frac{1}{(k\cdot v -i\varepsilon)^2}
+\frac{2\pi \delta(k\cdot v)}{\varepsilon}
=\frac{1}{(k\cdot v -i\varepsilon)^2}
+2\pi L\delta(k\cdot v)
$$
The result contains a well-defined derivative of a distribution, $\frac{1}{(x-i\epsilon)^2}\equiv -\frac{d}{dx}\frac{1}{x-i\epsilon}$.
It also  contains an infinite term, since $\frac{1}{\varepsilon}=\lim_{k\to 0}\int_0^\infty ds
e^{is(k+i\varepsilon)} = L$ is the infinite length of one semi-infinite Wilson line.
The length-$L$-dependent terms  appear as 
 $\sim\frac{1}{p^2+m^2}\times L\times $(momentum, $v$, $v'$-independent)  contributions which can be attributed to self-energies  of each of the two semi-infinite Wilson lines.  They can be absorbed into the normalizations of the dressed operators.  
 
 We assume that this has been done and we are left with the remainder of the first term and the second term, the renormalized correction
\begin{equation}\label{Dv' 4}
\begin{aligned}
\delta^{(2)}_{\rm WL}D_{vv'}(p)
&=-\frac{1}{2}e^2\mu^{4-D}\frac{1}{p^2+m^2}\int \frac{d^Dk}{(2\pi)^D}
\frac{1}{ k^2+m_\gamma^2}
 \frac{1}{(k\cdot v-i\varepsilon)^2}+(v\leftrightarrow v')
\\&
+e^2\mu^{4-D}\int \frac{d^Dk}{(2\pi)^D}
\frac{1}{[(p+k)^2+m^2][k^2+m_\gamma^2]}
\frac{v\cdot v'}{(k\cdot v -i\varepsilon)( k\cdot v'-i\varepsilon)}
 \end{aligned}
 \end{equation}
Now we are ready to study the infrared logarithm.
For this task, we can take the following procedure.   We operate $m_\gamma^2\frac{d}{dm_\gamma^2}$ on each of the terms, then we scale the integration variable $k\to km_\gamma$ in each term, and then we put $m_\gamma\to 0$.  The result is, for the infrared divergent part,
\begin{equation}\label{Dv' 5}
\begin{aligned}
D_{vv'}(p)=&\frac{1}{p^2+m^2}\frac{e^2}{2}\ln\frac{m_\gamma^2}{m^2}
 \int \frac{d^4k}{(2\pi)^4}
\frac{1}{ [k^2+1]^2}
\left(\frac{ v^\mu}{k\cdot v -i\varepsilon}
- \frac{ {v'}^\mu}{  k\cdot v'-i\varepsilon}\right)^2
+\ldots
\end{aligned}
\end{equation}
where the ellipses denote all contributions of order $e^2$ which remain finite as $m_\gamma$ is put to zero and also, all contributions of order  $e^4$ or higher.  The integrals are ultraviolet finite, so we have set $D=4$.

 We expand the squared quantity in  (\ref{Dv' 5}) to get
  \begin{equation}
\begin{aligned}
D_{vv'}(p)=&\frac{1}{p^2+m^2}e^2\ln\frac{m_\gamma^2}{m^2}
 \int_0^1d\alpha\int \frac{d^4k}{(2\pi)^4}
\frac{1}{ [k^2+1]^2}\left[\frac{1}{(k\cdot v -i\varepsilon)^2}
-\frac{v\cdot v'}{(k\cdot v_\alpha -i\varepsilon)^2 }\right]
+\ldots
\\& ~~~v_\alpha=\alpha v+(1-\alpha)v'
\end{aligned}
\end{equation}
where we have used the fact that the first term doesn't depend on $v$ and we have introduced a Feynman parameter to simplify the second term. Now we rewrite the distributions as
  \begin{equation}
\begin{aligned}
D_{vv'}(p)=&\frac{1}{p^2+m^2}e^2\ln\frac{m_\gamma^2}{m^2}
 \int_0^1d\alpha\int \frac{d^4k}{(2\pi)^4}
\frac{1}{ [k^2+1]^2}\left[(-v\cdot\partial_k)\frac{1}{k\cdot v -i\varepsilon}
-\frac{v\cdot v'}{v_\alpha^2}( -v_\alpha\cdot\partial_k)\frac{1}{k\cdot v_\alpha -i\varepsilon }\right]
+\ldots
\end{aligned}
\end{equation}
Upon integrating by parts and canceling the distribution altogether, we obtain
 \begin{equation}\label{Dv' 6}
\begin{aligned}
D_{vv'}(p)=&\frac{1}{p^2+m^2}e^2\ln\frac{m_\gamma^2}{m^2}
 \int_0^1d\alpha\int \frac{d^4k}{(2\pi)^4}
\frac{-4}{ [k^2+1]^3}\left[1
-\frac{v\cdot v'}{v_\alpha^2}\right]
+\ldots
\\=&-\frac{1}{p^2+m^2}\frac{e^2}{8\pi^2}\ln\frac{m_\gamma^2}{m^2}
\left[1
-\int_0^1d\alpha \frac{\cos\delta}{\alpha^2+(1-\alpha)^2+2\alpha(1-\alpha)\cos\delta }\right]
+\ldots
\\=&\frac{1}{p^2+m^2}\frac{e^2}{8\pi^2}
\left[\delta\cot\delta-1\right] \ln\frac{m_\gamma^2}{m^2}
+\ldots~~,~~~ v\cdot v'=\cos\delta
\end{aligned}
\end{equation}
It is also rather easy to see that higher orders will produce higher powers of this logarithmic term and, when summed up, this leading term exponentiates.  Moreover, there are no further infrared divergent contributions.  The infrared divergent part, to all orders in perturbation theory is thus given by 
  \begin{equation}\label{Dv' 7}
\begin{aligned}
D_{vv'}(p)=&\frac{1}{p^2+m^2}\left(\frac{m_\gamma^2}{m^2}\right)^{\frac{e^2}{8\pi^2} \left[\delta\cot\delta-1\right]}\biggl[1+~{\rm finite~as~}m_\gamma\to 0 ~\biggr]
\end{aligned}
\end{equation}
where the square bracket contains all contributions which remain finite as $m_\gamma\to 0$.
Note that, since the exponent in equation (\ref{Dv' 7}) is negative definite, the factor diverges as $m_\gamma\to 0$.  The situation is slightly different if we inverse Wick rotate, back to Minkowski space where the angle $\delta$ between Euclidean vectors $v$ and $v'$ is replaced by the rapidity angle $\delta \to -i\chi$ between forward directed time-like vectors so
that (\ref{Dv' 7}) is replaced by
 \begin{equation}\label{Dv' 8}
\begin{aligned}
D_{vv'}(p)=&\frac{-i}{p^2+m^2-i\varepsilon}\left(\frac{m_\gamma^2}{m^2}\right)^{\frac{e^2}{8\pi^2} \left[\chi \coth\chi-1\right]}\biggl[1+~{\rm finite~as~}m_\gamma\to 0 ~\biggr]
\end{aligned}
\end{equation}
where the exponent is positive definite.  This implies that our inner product
of state vectors goes to zero when $v\neq v'$ -- they become orthogonal.

We note the formal similarity of the exponent in this equation to the
one-loop cusp anomaly.  However, unlike the ultraviolet cusp
\cite{Polyakov:1980ca}, which has corrections at all orders in perturbation
theory, this infrared exponent is one-loop exact whenever the electrically
charged matter has a mass gap.  The coefficient of $\ln m_\gamma$ is
determined by photons with $k\to 0$, and the soft photon theorems
\cite{Low:1954kd,Low:1958sn,Burnett:1967km,Gribov:1966hs} control these
in two ways.  First, a soft photon couples to the Wilson lines through the
eikonal factor $v^\mu/(k\cdot v)$ by construction and, by Low's theorem,
to the nearly on-shell charged particle through the same factor with
$v^\mu$ replaced by the particle velocity, with corrections suppressed by
powers of $k/m$.  Second, a soft photon coupled to a charged line inside a
closed loop acquires no such singularity; together with Furry's theorem
this implies that the effective action for photons with $k\ll m$, obtained
by integrating out the charged matter, is quadratic up to terms of order
$F^4/m^4$, which are harmless by infrared power counting.  The soft
photons are therefore governed by a Gaussian functional of eikonal
currents, the one-loop exponent exponentiates, and the only quantity that
can appear in it is the soft photon propagator at $k\to0$, whose vacuum
polarization vanishes there when $e^2$ is normalized to the Thomson
cross-section.  (Gribov's qualification, that the soft expansion requires
$\omega\ll m^2/E$, is satisfied at fixed rapidity as $k\to0$.)  The formal
resemblance to the cusp anomalous dimension arises because the far
infrared photons cannot distinguish the semi-infinite Wilson line segments
from those which would meet at a cusp.

 This leaves the  $v\neq v'$ 2 point function infrared divergent and there is no cure for this divergence.  It is a physical manifestation of the fact that the long-ranged parts of the photon clouds in dressings with different $v$'s are sufficiently dissimilar that they have vanishing overlap as quantum states. (Note that the infrared divergence comes from the self-interaction of the Wilson lines only.)
 The infrared divergences will cancel identically only when $\delta=0$, that is, when $v=v'$. This is the photon cloud superselection rule. 
 
 \subsection{The superselected 2 point function $D_{vv}(p^2)$}
 
 It remains to study the infrared finite 2 point function with coincident $v$'s.  We return to the expressions in equation (\ref{Dv' 2}) and we set $v=v'$, 
  \begin{equation}\label{Dv' 9}
 \begin{aligned}
 D_{vv}(p)&=\frac{1}{p^2+m^2}+ \delta_1^{(2)}D_{vv}(p)
 + \delta_2^{(2)}D_{vv}(p)+ \delta_3^{(2)}D_{vv}(p)+\ldots
 \\
 \delta_1^{(2)}D_{vv}(p)&=-\frac{ \Pi_{ h}(p)}{[p^2+m^2]^2}
\\
 \delta^{(2)}_2D_{vv}(p)&
 =e^2\mu^{4-D}\int \frac{d^Dk}{(2\pi)^D}\left[\frac{1}{(p+k)^2+m^2}-\frac{1}{p^2+m^2}\right]
\frac{1}{ k^2+m_\gamma^2}
 \frac{1}{(k\cdot v -i\varepsilon)^2}
 \\ \delta^{(2)}_3 D_{vv}(p)&
 =e^2\frac{4p\cdot v}{{p}^2+m^2} \int \frac{d^4k}{(2\pi)^4} \frac{1
}{(p+k)^2+m^2}\frac{1}{k^2+m_\gamma^2}\frac{1}{k\cdot v-i\varepsilon}
\\&
+2e^2\mu^{4-D}\frac{1}{{p}^2+m^2} \int \frac{d^Dk}{(2\pi)^D} \frac{1
}{(p+k)^2+m^2}\frac{1}{k^2+m_\gamma^2}\end{aligned}
 \end{equation}

  The expression for $\delta^{(2)}_2D_{vv}(p)$ in equation (\ref{Dv' 9}) still contains the eikonal double pole which we again write as the derivative of the single pole distribution,
$\frac{1}{(k\cdot v-i\varepsilon)^{2}}=(-v\cdot\partial_k)\frac{1}{k\cdot v-i\varepsilon}$.  Then we integrate by parts. 
Using
$v\cdot(p+k)=p\cdot v+k\cdot v$ and
$\tfrac{k\cdot v}{k\cdot v-i\varepsilon}=1$, this produces 
\begin{equation}\label{delta2 final}
\begin{aligned}
\delta^{(2)}_2D_{vv}(p)
=&\;2p\cdot v\,\frac{\partial}{\partial m^{2}}\,
e^{2}\mu^{4-D}\!\int\frac{d^{D}k}{(2\pi)^{D}}
\frac{1}{[(p+k)^{2}+m^{2}]\,[k^{2}+m_\gamma^{2}]\,[k\cdot v-i\varepsilon]}
\\
&-2e^{2}\mu^{4-D}\!\int\frac{d^{D}k}{(2\pi)^{D}}
\frac{1}{[(p+k)^{2}+m^{2}]^{2}\,[k^{2}+m_\gamma^{2}]}
\\
&-2e^{2}\mu^{4-D}\!\int\frac{d^{D}k}{(2\pi)^{D}}
\left[\frac{1}{(p+k)^{2}+m^{2}}-\frac{1}{p^{2}+m^{2}}\right]
\frac{1}{[k^{2}+m_\gamma^{2}]^{2}}
\end{aligned}
\end{equation}

 Combining (\ref{delta2 final}) with
$\delta^{(2)}_3D_{vv}(p)$ from (\ref{Dv' 9}), we get
\begin{equation}\label{combined}
\begin{aligned}
\delta^{(2)}_2D_{vv}(p)&+\delta^{(2)}_3D_{vv}(p)
=2p\cdot v\left[\frac{2}{p^{2}+m^{2}}
+\frac{\partial}{\partial m^{2}}\right]
e^{2}\mu^{4-D}\!\int\frac{d^{D}k}{(2\pi)^{D}}
\frac{1}{[(p+k)^{2}+m^{2}]\,[k^{2}+m_\gamma^{2}]\,[k\cdot v-i\varepsilon]}
\\
&+2e^{2}\mu^{4-D}\!\int\frac{d^{D}k}{(2\pi)^{D}}
\biggl\{\frac{1}{p^{2}+m^{2}}\,\frac{1}{(p+k)^{2}+m^{2}}\,
\frac{1}{k^{2}+m_\gamma^{2}}
-
\frac{1}{[(p+k)^{2}+m^{2}]^{2}\,[k^{2}+m_\gamma^{2}]}
\\
&\qquad\qquad
-\frac{1}{(p+k)^{2}+m^{2}}\,\frac{1}{[k^{2}+m_\gamma^{2}]^{2}}
+\frac{1}{p^{2}+m^{2}}\,\frac{1}{[k^{2}+m_\gamma^{2}]^{2}}\biggr\}
\end{aligned}
\end{equation}

The one-loop integrals in the second, third and fourth lines of equation (\ref{combined}) are simple scalar integrals which turn out to be infrared finite and have a relatively simple limit as $m_\gamma\to 0$, which we evaluate to produce the expression
\begin{equation}\label{combined 1}
\begin{aligned}
&\delta^{(2)}_2D_{vv}(p)+\delta^{(2)}_3D_{vv}(p)
=2p\cdot v\left[\frac{2}{p^{2}+m^{2}}
+\frac{\partial}{\partial m^{2}}\right]
e^{2}\mu^{4-D}~I
\\ &
+\frac{1}{p^2+m^2}\frac{e^2}{4\pi^2}\left[\frac{1}{\epsilon}-\gamma_E+\ln\frac{4\pi\mu^2}{p^2+m^2}+1\right]
-\frac{e^2}{8\pi^2}\frac{1}{p^2}\ln\frac{p^2+m^2}{m^2}
\\ &
I~\equiv~\int \frac{d^4k}{(2\pi)^4} \frac{1}{(p+k)^2+m^2}\frac{1}{ k^2}\frac{1}{k\cdot v -i\varepsilon}
\end{aligned}
\end{equation}
The integral that remains to be taken is $I$ which comes from the first line of  equation (\ref{combined}).  Since the integral is ultraviolet and infrared finite, we have taken $D=4$ and we have set $m_\gamma=0$.
 We combine the denominators using Feynman parameters.  To deal with $\frac{1}{k\cdot v -i\varepsilon}$,  we have used 
 $$\frac{1}{A^2B}=\int_0^\infty d\beta\frac{2}{[A+\beta B]^3}$$  
 whence $i\epsilon$ defines the pole in the $\beta$-integrand. 
 \begin{align*}
 I&
~=~2\int_0^1d\alpha\int_0^\infty d\beta \int \frac{d^4k}{(2\pi)^4} \frac{1}{
[k^2+2\alpha kp+\beta(kv-i\varepsilon)+\alpha(p^2+m^2) ]^3
}
\end{align*}
We can then take the $k$ integral to obtain
\begin{align*}
I~=~\frac{1}{16\pi^2}\int_0^1d\alpha\int_0^\infty d\beta \frac{1}{
\alpha(p^2+m^2)  - \alpha^2p^2-\alpha\beta p\cdot v-i\beta\epsilon -\frac{1}{4}\beta^2
}
\end{align*}
The integral over $\beta$ is now elementary.  Taking $v$ to be a unit vector and labeling $p_\parallel\equiv p\cdot v$ and $p_\perp \equiv p-p_\parallel v$
so that $v\cdot p_\perp=0$, we find
\begin{equation}\label{artanh form}
I=-\frac{1}{8\pi^2}\int_0^1d\alpha
\frac{1}{\sqrt{\alpha(p^{2}+m^{2})-p_\perp^{2}\alpha^{2}}}
{\rm artanh}\frac{p_\parallel\alpha}{\sqrt{\alpha(p^{2}+m^{2})-p_\perp^{2}\alpha^{2}}}+\frac{i\phi}{8\pi|p|\sin\Theta }
\end{equation}
where the angles $\Theta$ and $\phi$ in the imaginary part are defined in equation (\ref{angles definition}) below.  To get the real part, consider
\begin{equation}\label{artanh form 1}
\begin{aligned}
&\Re I=-\frac{1}{8\pi^2}\int_0^1d\alpha\int_0^{p_\parallel}dw\frac{d}{dw}
\frac{1}{\sqrt{\alpha(p^{2}+m^{2})-p_\perp^{2}\alpha^{2}}}
{\rm artanh}\frac{w\alpha}{\sqrt{\alpha(p^{2}+m^{2})-p_\perp^{2}\alpha^{2}}}
\\ &
=-\frac{1}{8\pi^2}\int_0^1 d\alpha\int_0^{p_\parallel}dw\frac{1}{p^2+m^2-\alpha(p_\perp^2+w^2)}
=\frac{1}{8\pi^2}\int_0^{p_\parallel}dw\frac{1}{p_\perp^2+w^2}\ln\frac{p_\parallel^2+m^2-w^2}{p^2+m^2}
\end{aligned}
\end{equation}

We define the angular variables where $\Theta$ is the angle between $p$ and $v$ so that
\begin{equation}\label{angles definition}
\begin{aligned}
&
\cos\Theta=\frac{p_\parallel}{|p|},~~\sin\Theta=\frac{|p_\perp|}{|p|},~~0\leq\Theta\leq \pi
\\ & \cos\phi= \frac{\sqrt{p_\parallel^2+m^2}}{\sqrt{p^2+m^2}},~~
\sin\phi = \frac{|p_\perp|}{\sqrt{p^2+m^2}},~0\leq\phi\leq\frac{\pi}{2}
\end{aligned}
\end{equation}
and we change the $w$-integration variable
\begin{equation}\label{angles definition 1}
\begin{aligned}
w=|p_\perp|\tan\psi   ,~~0\leq w\leq p_\parallel\to 0\leq \psi\leq\arctan\frac{p_\parallel}{|p_\perp|}= \frac{\pi}{2}-\Theta
\end{aligned}
\end{equation}
whence the integral becomes the log-cosine integrals
\begin{align}
&\Re I=\frac{1}{8\pi^{2}|p|\sin\Theta}\int_0^{\frac{\pi}{2}-\Theta} d\psi
\left[\ln\cos(\psi+\phi)+\ln\cos(\phi-\psi)-2\ln\cos\psi \right]
\end{align}
The log-cosine integral is given by Clausen's function, for example
\begin{align*}
&\int_{0}^{\frac{\pi}{2}-\Theta}d\psi \ln(2\cos\psi)=\frac{1}{2}\int_{0}^{\frac{\pi}{2}-\Theta}d\psi [ \ln(1+e^{2i\psi})
+ \ln(1+e^{-2i\psi})]
 =-\sum_{n=1}^\infty \frac{(-1)^n}{n}\int_{0}^{\frac{\pi}{2}-\Theta}d\psi\cos[2n\psi]
\\ &
 =-\sum_{n=1}^\infty \frac{(-1)^n}{2n^2}\sin[n\pi -2n\Theta]
=\frac{1}{2}\sum_{n=1}^\infty\frac{ \sin(n(2\Theta)) }{n^2}
=\frac{1}{2}{\rm Cl}_{2}(2\Theta)
\end{align*}
where ${\rm Cl}_{2}(z)$ is defined by
\begin{align}\label{Clausen's function}
{\rm Cl}_{2}(z)~\equiv~\sum_{n=1}^\infty \frac{\sin(nz)}{n^2}
=-\int_{0}^{z}dt\ln\big|2\sin\tfrac t2\big|={\rm Im}{\rm Li}_{2}(e^{iz})
\end{align}
We finally have
\begin{equation}\label{I result}
\Re I=\frac{1}{16\pi^{2}|p|\sin\Theta}
\Big[{\rm Cl}_{2}\big(2\Theta+2\phi\big)+{\rm Cl}_{2}\big(2\Theta-2\phi\big)
-2{\rm Cl}_{2}\big(2\Theta\big)\Big]
\end{equation}
 and
\begin{equation}\label{I result corrected}
I=\frac{1}{16\pi^{2}|p|\sin\Theta}
\Bigl[{\rm Cl}_{2}\bigl(2\Theta+2\phi\bigr)+{\rm Cl}_{2}\bigl(2\Theta-2\phi\bigr)
-2\,{\rm Cl}_{2}\bigl(2\Theta\bigr)+2\pi i\,\phi\Bigr]
\end{equation}
where we have now included the imaginary part inside the bracket.  We shall also need
\begin{equation}\label{dIdm2}
\frac{d}{dm^{2}}I=\frac{1}{16\pi^{2}}\frac{1}{p^{2}+m^{2}}
\frac{1}{\sqrt{p_\parallel^{2}+m^{2}}}
\biggl[\ln\frac{\sqrt{p_\parallel^{2}+m^{2}}+p_\parallel}
{\sqrt{p_\parallel^{2}+m^{2}}-p_\parallel}~-~i\pi\biggr]
\end{equation}
Now, let us assemble the correction to the 2 point function as we have computed it,
\begin{equation}\label{assembly}
\begin{aligned}
D_{vv}(p)= \frac{1}{p^{2}+m^{2}}\biggl\{&\,1
-\frac{e^{2}}{8\pi^{2}}\biggl[
\Bigl(1-\frac{m^{2}}{p^{2}}\Bigr)\ln\frac{p^{2}+m^{2}}{m^{2}}
-\ln\frac{\mu^{2}}{m^{2}}-2\biggr]
\\&
+\frac{e^{2}}{4\pi^{2}}\left[\frac{1}{\epsilon}-\gamma_{E}
+\ln\frac{4\pi\mu^{2}}{p^{2}+m^{2}}+1\right]
-\frac{e^{2}}{8\pi^{2}}\,\frac{p^{2}+m^{2}}{p^{2}}\,
\ln\frac{p^{2}+m^{2}}{m^{2}}
\\&
+\frac{e^{2}}{8\pi^{2}}\,\frac{p_\parallel}{\sqrt{p_\parallel^{2}+m^{2}}}
\biggl[\,\ln\frac{\sqrt{p_\parallel^{2}+m^{2}}+p_\parallel}
{\sqrt{p_\parallel^{2}+m^{2}}-p_\parallel}-i\pi\biggr]
\\&
+\frac{e^{2}}{4\pi^{2}}\,\frac{p_\parallel}{|p_\perp|}
\Bigl[{\rm Cl}_{2}\bigl(2\Theta+2\phi\bigr)
+{\rm Cl}_{2}\bigl(2\Theta-2\phi\bigr)
-2\,{\rm Cl}_{2}\bigl(2\Theta\bigr)+2\pi i\phi\Bigr]
\\&
~~+~~\mathcal O(e^4)~~\biggr\}
\end{aligned}
\end{equation}
This equation (\ref{assembly}) is the central result of this section, the 2 point function of the Wilson-line-dressed charged scalar
fields to order $e^2$.
The first order $e^2$ bracket in the first line is the hybrid-subtracted self-energy correction of the undressed particle.
The second line collects the infrared finite loop integrals of
(\ref{combined}); the
third and fourth lines are
$2p\cdot v\,e^{2}\,\partial I/\partial m^{2}$ and
$4p\cdot v\,e^{2}\,I/(p^{2}+m^{2})$ respectively, with $I$ from
(\ref{I result corrected}).  The imaginary parts are odd under
$p_\parallel\to-p_\parallel$, consistent with the reality of the
position-space correlator, $D_{vv}(p)^{*}=D_{vv}(-p)$. They arise from
the photon modes with $k\cdot v=0$, which are static in the frame of the
line, and they will play an essential role in the mass-shell limit
below.  The single remaining ultraviolet pole is removed by the
multiplicative renormalization of the dressed operators.

\subsection{The mass-shell limit and the infraparticle exponent}
What remains is to examine our result appearing in equation (\ref{assembly}) above 
for the residue obstruction.  That is, we need to evaluate the coefficient of
$\frac{1}{p^2+m^2}\ln(p^{2}+m^{2})$ that will appear in (\ref{assembly}) as the limit
$p^{2}\to-m^{2}$ is taken. 

 In the vicinity of the mass
shell the angles $\Theta$ and $\phi$ become complex.
Rather than
continuing the Clausen functions directly it is simpler to extract the
mass-shell logarithm from an earlier integral form of $\Re I$ which
appeared in the last equality in equation (\ref{artanh form 1})
\begin{equation}
\Re I =-\frac{1}{8\pi^2|p_\perp|}\arctan\frac{p_\parallel}{|p_\perp|}\ln\frac{p^2+m^2}{m^2}
+\frac{1}{8\pi^2}\int_0^{p_\parallel}dw\frac{1}{p_\perp^2+w^2}\ln\frac{p_\parallel^2+m^2-w^2}{m^2}
\end{equation}
The entire mass shell logarithm sits in the first term. Its coefficient is
$\arctan\frac{p_\parallel}{|p_\perp|}=\frac{\pi}{2}-\Theta$.  Also, when we analytically continue to the mass shell, $p_\perp^2=p^2-p_\parallel^2\to -(p_\parallel^2+m^2)$ so that $|p_\perp|\to i\sqrt{p_\parallel^2+m^2}$ and
$$
\arctan\frac{p_\parallel}{|p_\perp|}~\to~-i\,{\rm artanh}\frac{p_\parallel}{\sqrt{p_\parallel^2+m^2}}
$$
so that
$$
-\frac{1}{8\pi^2|p_\perp|}\arctan\frac{p_\parallel}{|p_\perp|}
~\to~\frac{1}{16\pi^2\sqrt{p_\parallel^2+m^2}}\ln\frac{\sqrt{p_\parallel^2+m^2} +p_\parallel}{\sqrt{p_\parallel^2+m^2}-p_\parallel}
$$
Also, for the imaginary part,
$$
i~\Im I~ =~\frac{i\phi}{8\pi|p_\perp|} = \frac{i}{8\pi|p_\perp|}\arcsin\frac{|p_\perp|}{\sqrt{p^2+m^2}}\to  \frac{i}{8\pi|p_\perp|}\,i\,{\rm arcsinh}\frac{\sqrt{p_\parallel^2+m^2}}{\sqrt{p^2+m^2}}
$$
$$
=\frac{-i\pi}{16\pi^2 \sqrt{p_\parallel^2+m^2}}\ln\frac{p^2+m^2}{m^2}
+\ldots
$$
where the ellipses in the last equality denote parts which are not singular as $p^2\to-m^2$.
In summary
\begin{equation}\label{I near shell}
I=\frac{1}{16\pi^{2}\sqrt{p_\parallel^{2}+m^{2}}}
\biggl[\ln\frac{\sqrt{p_\parallel^{2}+m^{2}}+p_\parallel}
{\sqrt{p_\parallel^{2}+m^{2}}-p_\parallel}-i\pi\biggr]
\,\ln(p^{2}+m^{2})+~({\rm regular~at~}p^{2}=-m^{2})
\end{equation}

On the mass shell we parameterize the projection of the momentum on the
Wilson line direction by the rapidity $\zeta$,
\begin{equation}\label{rapidity}
p\cdot v=im\cosh\zeta~,~~
\sqrt{p_\parallel^{2}+m^{2}}=im\sinh\zeta
\end{equation}
which is the Euclidean continuation of $v\cdot p=-m\cosh\zeta$ for
forward directed time-like Minkowski vectors.  Then
$$\ln\frac{\sqrt{p_\parallel^{2}+m^{2}}+p_\parallel}
{\sqrt{p_\parallel^{2}+m^{2}}-p_\parallel}
=\ln(-e^{2\zeta})=2\zeta+i\pi$$ the explicit $-i\pi$ in
(\ref{I near shell}) cancels, and
\begin{equation}\label{I on shell}
I=-\frac{i\zeta}{8\pi^{2}m\sinh\zeta}~\ln(p^{2}+m^{2})
+({\rm regular})
\end{equation}

The imaginary part of $I$, which descends from the static photon modes
$k\cdot v=0$, is essential here.  Without it, the continuation would be
off by a spurious $i\pi$.

Inserting (\ref{I on shell}) into the last line of
(\ref{assembly}) with $4p\cdot v=4im\cosh\zeta$ gives the
mass-shell logarithm
$+\frac{e^{2}}{8\pi^{2}}4\zeta\coth\zeta\ln(p^{2}+m^{2})$ inside the
braces.  The self-energy bracket contributes
$-\frac{e^{2}}{8\pi^{2}}\cdot2 \ln(p^{2}+m^{2})$, since
$1-m^{2}/p^{2}\to2$ on the mass shell; the infrared finite loop integrals
contribute $-\frac{e^{2}}{8\pi^{2}}\cdot 2 \ln(p^{2}+m^{2})$.
The rest of the terms are regular on the mass shell and do not contribute. 

Altogether we obtain
\begin{equation}\label{near shell D}
D_{vv}(p)= \frac{1}{p^{2}+m^{2}}\biggl\{1
+\frac{e^{2}}{2\pi^{2}}\bigl(\zeta\coth\zeta-1\bigr)
\ln\frac{p^{2}+m^{2}}{m^{2}}
+\ldots\biggr\}
\end{equation}
where the ellipses denote terms which have a finite limit on
the mass shell.  

As in the infrared logarithms of the previous
subsection, the higher orders exponentiate, so that
\begin{equation}\label{infraparticle exponent}
\begin{aligned}
&D_{vv}(p)~=~\left[\frac{1}{p^{2}+m^{2}}\right]^{1-\gamma}
~+~{\rm nonsingular~at~}p^2\to-m^2
\\&
\gamma=\frac{e^{2}}{2\pi^{2}}\bigl(\zeta\coth\zeta-1\bigr)
\end{aligned}
\end{equation}

This is the residue obstruction for the Wilson-line dressed operators.
For generic orientation of the dressing, as long as the dressing lines are parallel, 
the mass-shell pole is replaced by the branch point of equation (\ref{infraparticle exponent}), with the
computable, positive semi-definite exponent $\gamma$.

The exponent in
(\ref{infraparticle exponent}) is precisely the coefficient of
$\ln(m_\gamma/m)$ that appears in the on-shell residue if we take limits
in the other order, we keep $m_\gamma$ nonzero and we put $p^2\to-m^2$.
In that scheme the residue is
proportional to $(m_\gamma/m)^{\gamma(\zeta)}$.  The trade of
$\ln(p^{2}+m^{2})$ at $m_\gamma=0$ for $\ln m_\gamma$ at
$p^{2}=-m^{2}$ is the same non-commutativity of limits as we encountered for
the undressed self-energy in section \ref{self energy}.

The exponent $\gamma(\zeta)$ vanishes
quadratically at the tuned point, $\zeta=0$, 
$$
\gamma=\frac{e^{2}}{6\pi^{2}}\zeta^{2}+O(\zeta^{4})
$$
When the
orientation of the dressing is fine tuned to the mass-shell momentum,
$v^{\mu}=\frac{1}{m}p^{\mu}$, that is, when $\zeta=0$, the mass-shell logarithm
cancels between the self-interaction of the dressing, the
interaction of the dressing with the particle, and the self-energy of
the particle.

\section{The massive photon and the infrared window}\label{window}

In this section we examine the case where we retain a small but finite infrared 
cutoff in the form of a small photon mass.  In that case, the orthogonality and the anomalous infraparticle scaling which we found in the previous section go away and they are replaced by a milder damping, in the case of photon cloud orthogonality, and the more conventional scenario of a pole at the physical mass, $p^2=-m^2$  of the charged particle in the two point function and a cut singularity beginning at the charged particle plus massive photon threshold $-p^2\geq (m+m_\gamma)^2$. 

The goal of this section is to show that, because the photon is exceedingly light, $\frac{m_\gamma}{m}<<1$, there is a nontrivial window of coarse grained energies or, in real space,  proper times where the infraparticle scaling law still holds.  

After creation of the dressed charged particle, the 2 point function achieves its asymptotic limit in a very short proper time, $t\sim \frac{1}{m}$, with $ \frac{1}{m}$ being the Compton time of the massive charged particle.  
Then,  in the scenario that we are interested in, with $\frac{m_\gamma}{m}<<1$, the momentum space gap between the pole and the onset of the cut is resolvable only with sufficient observer time, $t>\frac{1}{m_\gamma}$.  At times shorter than that, the pole and cut structure is indistinguishable from the infraparticle and the infraparticle scaling law (\ref{infraparicle scaling}) which we copy here
\begin{equation}\label{infraparicle scaling 1}
\begin{aligned}
&{\rm In~}\frac{1}{m}\lesssim t\lesssim \frac{1}{m_\gamma}
~:~~~~\mathcal A_{\rm infraparticle}(t)\sim \frac{ e^{-imt}}{t^{\frac{3}{2}+\gamma(\zeta)}}
\\
&\gamma(\zeta)=\frac{e^{2}}{2\pi^{2}}\bigl(\zeta\coth\zeta-1\bigr)
\\&
t=\sqrt{-x^\mu x_\mu}~,~~
\cosh\zeta =- \frac{v^\mu x_\mu}{\sqrt{-x^\mu x_\mu}}
\end{aligned}
\end{equation}
still holds.   We call this interval the infrared window.

\begin{figure}[h]
\centering
\includegraphics[width=0.85\textwidth]{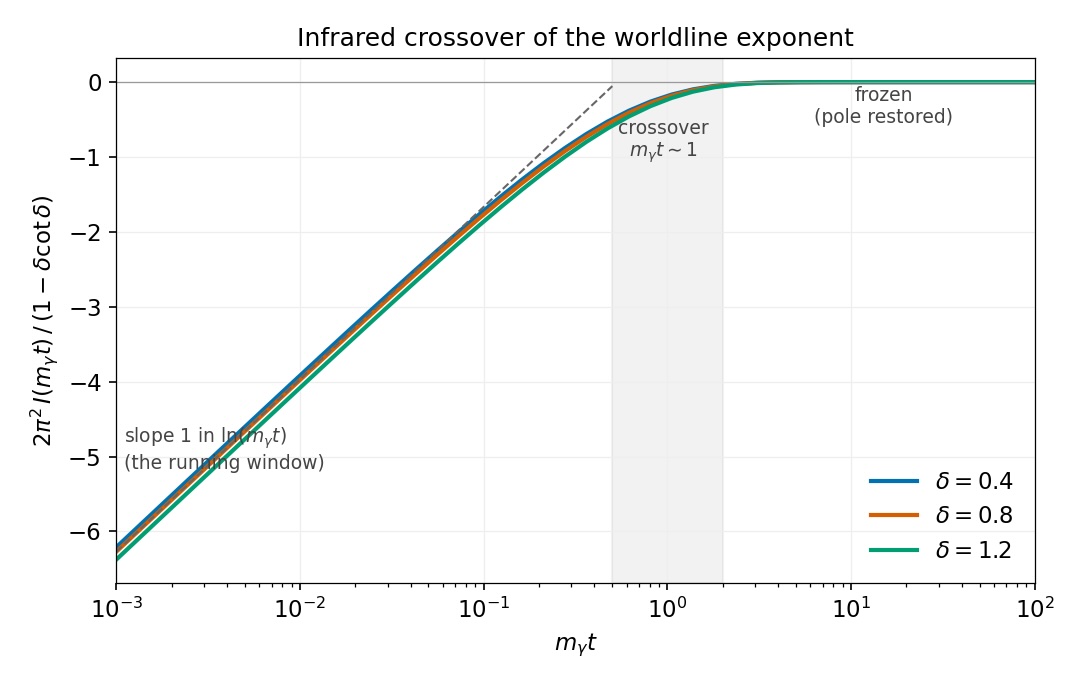}
\caption{The infrared crossover of the worldline exponent. Plotted is the
$t$-dependent part of the photon exchange integral, normalized as
$2\pi^{2}I(m_\gamma t)/(1-\delta\cot\delta)$, for three Euclidean angles
$\delta$.  In the window $m_\gamma t \lesssim 1$ the curves collapse onto a line
of unit slope in $\ln(m_\gamma t)$ -- the running logarithm with
coefficient $\frac{e^{2}}{2\pi^{2}}(1-\delta\cot\delta)$ -- and for
$m_\gamma t\gtrsim1$ they freeze: the $t$-dependence stops and the pole
is restored.}
\label{fig:crossover}
\end{figure}

After the window, at $t\sim\frac{1}{m_\gamma}$, the Compton time of the photon, the power law behaviour freezes. The factor $t^{-\gamma(\zeta)}$ is replaced by $\left(\frac{m_\gamma}{m}\right)^{\gamma(\zeta)}$.
One can see from the numerical plot in figure \ref{fig:crossover}, the width of the crossover is relatively narrow, about 2 decades of $\frac{0.1}{m_\gamma}\lesssim t\lesssim \frac{10}{m_\gamma}$. 
 
To find the infrared window, we should return to the perturbative computation of $D_{vv}(p^2)$ at order $e^2$, the contributions summarized  in equation
 (\ref{Dv' 9}) and plug in the more explicit forms of equation (\ref{combined}) and evaluate them with $m_\gamma$ nonzero. 
 We would need the Fourier transform of the momentum space 2 point function that is computed
 there, in the limit of large proper time, $t$.  
 
 In the following, we will use the alternative of a worldline functional integral approach to computing the 2 point function.  Worldline techniques are well developed and we refer the reader to Schubert's review \cite{schubert2001} for the details.  It will give us a direct approach to the large proper time limit of the 2 point function. 
 
Consider the Euclidean worldline representation of the Wilson line dressed 2 point function 
\begin{equation}\label{wlpi}
\begin{aligned}
&\mathcal A(t)~=~\langle ~
~\int_0^\infty dT \int[dy]
e^{ -\int_0^1 d\tau
\left[\frac{\dot y^2(\tau)}{4T}+m^2T\right]
+ie\int A } ~\rangle
\\
&
y^\mu(1)=x^\mu,~y^\mu(0)=0
\\
&
\int A = \int_{-\infty}^0 ds v\cdot A(vs)+\int_0^1d\tau\dot y(\tau) \cdot A(y(\tau))+\int_0^\infty
ds v\cdot A(sv+x)
\end{aligned}
\end{equation}
The functional integral is over the embedding functions $y^\mu(\tau)$ of a worldline into D dimensional Euclidean space.  The Dirichlet boundary condition on these functions encode the fact that they must begin at 
$y=0$ and end at $y=x$.  The coupling to the photon in $\int A$ contains the Wilson line integral along the particle trajectory $y(\tau)$ as well as the integral along the semi-infinite straight-line dressings characterized by the vector $v$. 
The outer bracket $\langle~ ...~ \rangle$ denotes whatever further averaging, particularly over $A$, would be needed to turn this into the full Euclidean 2 point function of the quantum field theory.

It is convenient to change the worldline functional integration variable $y(\tau)$ as 
$$
y(\tau)=x^\mu \tau+q^\mu (\tau)~,~~q^\mu(0)=0=q^\mu (1)
$$
When $e=0$ the functional integration over $y(\tau)$ would produce the free particle propagator. The functional integral is 
$$
\int[dq]
e^{ -\int_0^1 d\tau
 \frac{\dot q^2(\tau)}{4T}
 }~=~\frac{1}{(4\pi T)^{\frac{D}{2}}}
 $$
In that case, we can  re-write (\ref{wlpi})  as
\begin{equation}\label{wlpi 1}
\begin{aligned}
&\mathcal A(t)~=~\langle ~
~\int_0^\infty \frac{dT}{(4\pi T)^{\frac{D}{2}}} e^{-\frac{t^2}{4T}-m^2T} ~\frac{\int[dq]
e^{ -\int_0^1 d\tau
 \frac{\dot q^2(\tau)}{4T}
+ie\int A }  }{\int[dq]
e^{ -\int_0^1 d\tau
 \frac{\dot q^2(\tau)}{4T}
 }}~\rangle
\\
&
q^\mu(1)~=~0~=~q^\mu(0)~,~~t^2~\equiv x^2
\\
&
\int A = \int_{-\infty}^0 ds v\cdot A(vs)+\int_0^1d\tau (x +\dot q(\tau) )\cdot A(x\tau+q(\tau))+\int_0^\infty
ds v\cdot A(sv+x)
\end{aligned}
\end{equation}
In this expression, $T$ has dimensions of inverse mass squared.  It is convenient to redefine it so that it is dimensionless, and at the same time, so that its saddle point value in the limit where $mt$ is large will be $T\sim 1$.  This is accomplished by
the rescaling
\begin{equation}\label{rescale T}
T\to \frac{t}{2m}T
\end{equation}
Then, the kinetic term in the action for $q$ takes on the form $\dot q^2 m/2Tt$.  We rescale $q$ so
that its saddle point is also independent of $m$ and $t$,
\begin{equation}\label{rescale q}
q(\tau)\to \sqrt{\frac{t}{m}}q(\tau)
\end{equation}
The Jacobian for rescaling $q(\tau)$ cancels between the numerator and the denominator in the 
ratio of functional integrals.  

The result is
\begin{equation}\label{wlpi 2}
\begin{aligned}
&\mathcal A(t)~=~\langle ~\left(\frac{t}{2m}\right)^{1-\frac{D}{2}}
~\int_0^\infty \frac{dT}{(4\pi T)^{\frac{D}{2}}} e^{-\frac{tm}{2}(T+1/T)}~\frac{\int[dq]
e^{ -\int_0^1 d\tau
 \frac{\dot q^2(\tau)}{2T}
+ie\int A }  }{\int[dq]
e^{ -\int_0^1 d\tau
 \frac{\dot q^2(\tau)}{2T} }
 }~\rangle
\\
&
q^\mu(1)~=~0~=~q^\mu(0)
\\
&
\int A = \int_{-\infty}^0 ds v\cdot A(vs)+\int_0^1d\tau t(\hat x +\frac{\dot q(\tau)}{\sqrt{mt} })\cdot
A(t(\hat x\tau+\frac{ q(\tau)}{\sqrt{mt}} ))+\int_0^\infty
ds v\cdot A(sv+x)
\end{aligned}
\end{equation}
Now, we are prepared to take the large $mt$ limit where we expand the $\int A$ as 
\begin{equation}
\begin{aligned}
&\int_0^1d\tau t(\hat x +\frac{\dot q(\tau)}{\sqrt{mt}} )\cdot
A(t(\hat x\tau+\frac{ q(\tau)}{\sqrt{mt}} ))=
\int_0^1d\tau x \cdot
A(x\tau)
+\frac{t^2}{\sqrt{mt}}\int_0^1d\tau q^\mu \hat x^\nu F_{\mu\nu}(x\tau)
\\ &~~~~~
+\frac{t^2}{2mt}\int_0^1d\tau\left[ \dot q^{\mu}q^{\nu}F_{\nu\mu}(x\tau)
+ q^{\nu}q^{\rho}x^{\mu}\partial_{\rho}F_{\nu\mu}(x\tau)\right]+\mathcal O \left(\frac{1}{\sqrt{mt}}\right)^3
\\&
 F_{\mu\nu}(x\tau) = \nabla_\mu A_\nu(x\tau)-\nabla_\nu A_\mu(x\tau)
\end{aligned}
\end{equation}
The appearance of $t^2$ in the coefficients of later terms will be canceled by the scaling of
the contractions of the photon fields which occur in the integrands.  
 
 A double expansion in $e^2$ and $1/mt$ of the 2 point function then contains
\begin{equation}\label{double expansion}
\begin{aligned}
\mathcal A(t)~=~& \left(\frac{t}{2m}\right)^{1-\frac{D}{2}}
~\int_0^\infty \frac{dT}{(4\pi T)^{\frac{D}{2}}} e^{-\frac{mt}{2}(T+1/T)}~\times\\ & \times
\biggl\{
 1
-\frac{e^2}{2}\Big\langle \Big(  \int A_{\rm cl}\Big)^2\Big\rangle
\\ &
 -\frac{e^2}{2}\frac{t^2}{mt}\int_0^1  d\tau d\tau'  x^\mu x^\rho
 \langle q^\nu (\tau)q^\sigma (\tau')\rangle
\langle~F_{\mu\nu} \big(x\tau\big)~F_{\rho\sigma} \big(x\tau'\big)\rangle
\\ &
-e^2\frac{t^2}{2mt}\int_0^1  d\tau \,
\langle  \dot q^\mu (\tau)q^\nu (\tau)\rangle
\Big\langle~F_{\nu\mu} \big(x\tau\big)~\int A_{\rm cl} \Big\rangle
\\ &
-e^2\frac{t^2}{2mt}\int_0^1  d\tau \, x^\mu
 \langle q^\nu (\tau)q^\rho (\tau) \rangle \Big\langle~\partial_\rho F_{\nu\mu} \big(x\tau\big)~\int A_{\rm cl} \Big\rangle
~+~\ldots~\biggr\}
\\ &
\int A_{\rm cl}~\equiv~\int_{-\infty}^0 ds\, v\cdot A(vs)+\int_0^1d\tau\, x \cdot
A(x\tau)+\int_0^\infty ds\, v\cdot A(sv+x)
\\ &
 \langle q^\nu (\tau)q^\sigma (\tau')\rangle
~\equiv~\frac{
\int[dq] e^{
-\int_0^1 d\tau
 \frac{\dot q^2(\tau)}{2T}
 }
 q^\nu (\tau)q^\sigma (\tau')
 }{
 \int[dq]
e^{ 
-\int_0^1 d\tau
 \frac{\dot q^2(\tau)}{2T} 
 } }
\end{aligned}
\end{equation}
where the ellipses stand for all contributions of order $e^4$ and higher
and order $1/(mt)^2$ and higher.  The cross term between the classical
line and the first-order fluctuation vanishes since $\langle q\rangle=0$.
The second line in (\ref{double expansion}) is the correction from
coupling to the classical Wilson line, consisting of the two semi-infinite
dressing segments and the segment following the classical trajectory of
the particle. The third, fourth and fifth lines summarize the
$\mathcal O(1/mt)$ corrections.  In dimensional regularization these are
finite, carry no pole in $\epsilon$ and therefore no $\ln t$; the
anomalous power can only arise from the second line.

If we set $e^2=0$ in (\ref{double expansion}), the remaining expression  produces the free field theory 2 point function which itself has an asymptotic expansion in $1/(mt)$, 
\begin{equation}\label{double expansion 1}
\begin{aligned}
\mathcal A_{\rm particle}(t)~&=
~\frac{2m}{t}
\int_0^\infty \frac{dT}{(4\pi T)^{\frac{D}{2}}} e^{-\frac{mt}{2}(T+1/T)}~=~
\frac{mK_1(mt)}{4\pi^2t}
\\
&=\frac{m^{\frac{1}{2}} }{2(2\pi)^{\frac{3}{2}}}
\frac{e^{-mt}}{t^{\frac{3}{2}}}\left[1+\frac{3}{8mt}-\frac{15}{128(mt)^2}+\ldots\right]
\end{aligned}\end{equation}
We are interested in corrections to the exponent of the $1/t^{\frac{3}{2}}$ power law.  
These corrections  will appear as terms $\sim \ln(t)$  in the expansion in equation (\ref{double expansion}).
They can only come from the first, order $-\frac{e^2}{2}<\int A\int A>$ term. 
We can drop the inverse powers of $mt$ in the expansion (\ref{double expansion}). 

It therefore remains to study  
\begin{align*}
&-\frac{e^2}{2}<\int A\int A>=-\frac{e^2}{2}\mu^{4-D}\langle \biggl(  \int_{-\infty}^0 ds v\cdot A(vs)+\int_0^1d\tau x \cdot
A(x\tau)+\int_0^\infty
ds v\cdot A(sv+x)\biggr)^2\rangle
\\
&=
e^2\mu^{4-D}\int \frac{d^Dk}{(2\pi)^D}\frac{1}{k^2+m_\gamma^2}(e^{ikx}  -1)
\biggl[ \frac{1}{(k\cdot v+i\varepsilon)^2} + \frac{1}{(k\cdot \hat x+i\varepsilon)^2}
-\frac{2v\cdot x}{(k\cdot v+i\varepsilon)(k\cdot x+i\varepsilon)}\biggr]
\\
&~~~~~~~~~~~~~~~
-e^2 \mu^{4-D}\int \frac{d^Dk}{(2\pi)^D}\frac{1}{k^2+m_\gamma^2}
\frac{1}{k\cdot v+i\varepsilon}\biggl[ \frac{1}{k\cdot v-i\varepsilon} - \frac{1}{k\cdot v+i\varepsilon} \biggr]
\end{align*}
The last, singular, $t$-independent term is the Wilson line energy density times its length, $\sim m_\gamma L$, where $L$ is the infinite length of the Wilson line segments and we drop it, assuming that it is absorbed by the normalization of the Wilson line operators.

In the remaining integral there are two terms, 
$$
-\frac{e^2}{2}<\int A\int A>=-\frac{e^2}{2}<\int A\int A>_{\rm exp}-\frac{e^2}{2}<\int A\int A>_{\rm non-exp}
$$
where $-\frac{e^2}{2}<\int A\int A>_{\rm exp}$ contains the exponential $e^{ikx}$ and $-\frac{e^2}{2}<\int A\int A>_{\rm non-exp}-$ does not. 

The non-exponential terms
\begin{align*}
&-\frac{e^2}{2}<\int A\int A>_{\rm non-exp}=
\\&
-e^2\mu^{4-D}\int \frac{d^Dk}{(2\pi)^D}\frac{1}{k^2+m_\gamma^2} 
\biggl[ \frac{1}{(k\cdot v+i\varepsilon)^2} + \frac{1}{(k\cdot \hat x+i\varepsilon)^2}
-\frac{2v\cdot x}{(k\cdot v+i\varepsilon)(k\cdot x+i\varepsilon)}\biggr]
\end{align*}
are integrals of the type which we have
already done in section \ref{wilson lines}.

For any unit vector $v$ one has the identity
\begin{align}
\int\frac{d^Dk}{(2\pi)^D}\frac{1}{k^2+m_\gamma^2}
\frac{1}{(k\cdot v+i\varepsilon)^2}
=-2\int\frac{d^Dk}{(2\pi)^D}\frac{1}{(k^2+m_\gamma^2)^2}
\label{ibp-identity}
\end{align}
This was obtained by writing $1/(k\cdot u+i\varepsilon)^2$ as minus the derivative of
$1/(k\cdot u+i\varepsilon)$ with respect to $k_\parallel=k\cdot u$ and
integrating by parts. The boundary term vanishes in dimensional
regularization, and the even part of $-2k_\parallel/(k_\parallel+i\varepsilon)$
is $-2$. Together with the standard tadpole Feynman integral
\begin{align}
\mu^{2\epsilon}\int\frac{d^Dk}{(2\pi)^D}\frac{1}{(k^2+m_\gamma^2)^2}
=\frac{1}{16\pi^2}\left[\frac{1}{\epsilon}-\gamma_E+\ln 4\pi
+\ln\frac{\mu^2}{m_\gamma^2}\right]
\label{tadpole}
\end{align}
each of the two squared eikonal factors contributes
$+\frac{e^2}{8\pi^2}$ times the bracket of (\ref{tadpole}).
For the cross term, combine the two eikonal denominators with a Feynman
parameter,
\begin{align}
\frac{1}{(k\cdot\hat x)(k\cdot v)}
=\int_0^1 d\alpha\,
\frac{1}{\bigl[k\cdot(\alpha\hat x+(1-\alpha)v)\bigr]^{2}}\,,
\label{feynman-combine}
\end{align}
apply (\ref{ibp-identity}) again---which for a non-unit combination picks up
a factor $1/(\alpha\hat x+(1-\alpha)v)^2$---and use
\begin{align}
\int_0^1 d\alpha
\frac{\cos\delta}{\alpha^2+(1-\alpha)^2+2\alpha(1-\alpha)\cos\delta}
=\cos\delta\cdot\frac{\delta}{\sin\delta}
=\delta\cot\delta
\label{alpha-integral}
\end{align}
where the denominator is
$\cos^2\frac{\delta}{2}+4\bigl(\alpha-\tfrac12\bigr)^2\sin^2\frac{\delta}{2}$
and the arctangent evaluates at exactly $\pm\delta/2$ at the endpoints.
The cross term therefore contributes $-\frac{e^2}{4\pi^2}\delta\cot\delta$
times the bracket, and the total is
\begin{align}
&-\frac{e^2}{2}<\int A\int A>_{\rm non-exp}=
\frac{e^2}{4\pi^2}
\left[\frac{1}{\epsilon}-\gamma_E+\ln 4\pi
+\ln\frac{\mu^2}{m_\gamma^2}\right]
\bigl[\,1-\delta\cot\delta\,\bigr]
\label{frozen-result}
\end{align}
This piece is independent of $t$. It renormalizes the residue (the overall
normalization of the two-point function) and does not modify the
$t^{-3/2}$ prefactor.   It is what the photon mediated interaction approaches,  in the very long proper time limit $t\gtrsim \frac{1}{m_\gamma}$ and, in that limit, the power law is the unmodified $1/t^{\frac{3}{2}}$. 

In the integral with the exponential, $-\frac{e^2}{2}<\int A\int A>_{\rm exp}$, we choose the orientation of the k-frame so that $x^\mu=t\delta^{\mu0}$.
Then we integrate using Cauchy's theorem by completing the $k_0$ contour in the upper half-plane. (We assume that $v\cdot \hat x\geq0$. ) 
\begin{align*}
&-\frac{e^2}{2}<\int A\int A>_{\rm exp}=
\\&
=
e^2\int \frac{d^{3}k}{(2\pi)^{3}}\frac{e^{-\sqrt{k^2+m_\gamma^2}~t}}{2\sqrt{k^2+m_\gamma^2}}
\biggl[ \frac{1}{(k\cdot v+i\varepsilon)^2} - \frac{1}{(k^2+m_\gamma^2)}
-\frac{2v\cdot \hat x}{(k\cdot v+i\varepsilon)(i\sqrt{k^2+m_\gamma^2})}\biggr]_{k_0=i\sqrt{k^2+m_\gamma^2}}
\end{align*}
The integral is ultraviolet finite and we have put $D=4$. 
Then we change variables, $k\to m_\gamma k $ to get
\begin{align*}
&-\frac{e^2}{2}<\int A\int A>_{\rm exp}=
\\&
=
\frac{e^2}{4\pi^2}\int_0^\infty k^2dk\frac{e^{-\sqrt{k^2+1}m_\gamma t}}{2\sqrt{k^2+1}}
\int_{-1}^1d\lambda \biggl[ -\frac{1}{(\cos\delta \sqrt{k^2+1}-ik\sin\delta \lambda)^2} - \frac{1}{(k^2+1)}
\\
&
~~~~~~~~~~~~~~~~~~~~~~~~~~~
+\frac{2\cos\delta }{(\cos\delta \sqrt{k^2+1}-ik\sin\delta\lambda)(\sqrt{k^2+1})}\biggr] 
\\
&=
-\frac{e^2}{4\pi^2}\int_0^\infty k^2dk\frac{e^{-\sqrt{k^2+1}m_\gamma t}}{\sqrt{k^2+1}}
\biggl[ \frac{1}{\cos^2\delta +k^2} + \frac{1}{(k^2+1)}
-\frac{2\cot\delta}{k\sqrt{k^2+1}}\arctan\frac{k\tan\delta}{\sqrt{k^2+1} } \biggr] 
\end{align*}
where we have separated the radial and angle integrals and taken the angular integral. 
Now, we change variables $k=\sinh\eta$ to get
\begin{align*}
&-\frac{e^2}{2}<\int A\int A>_{\rm exp}=
\\&
=
-\frac{e^2}{4\pi^2}\int_0^\infty d\eta e^{-m_\gamma t \cosh\eta }
\biggl[ \frac{\sinh^2\eta}{\cosh^2\eta-\sin^2\delta} + \tanh^2\eta
-2\cot\delta\tanh\eta ~\arctan[\tanh\eta \tan\delta]\biggr] 
\end{align*}
As $\eta\to\infty$ the bracket tends to
\begin{align}
&\biggl[ \frac{\sinh^2\eta}{\cosh^2\eta-\sin^2\delta} + \tanh^2\eta
-2\cot\delta\tanh\eta ~\arctan[\tanh\eta \tan\delta]\biggr] 
\\&~\longrightarrow~\left[
1+1-2\cot\delta\cdot\arctan(\tan\delta)\right]
=2-2\delta\cot\delta
\label{bracket-tail}
\end{align}
using $\arctan(\tan\delta)=\delta$ for $0\le\delta<\pi/2$, the same
restriction $v\cdot\hat x>0$ already assumed in closing the contour.
Adding and subtracting this constant inside the integral, and using
\begin{align}
\int_0^\infty d\eta\;e^{-m_\gamma t\cosh\eta}=K_0(m_\gamma t)\,,
\label{k0-def}
\end{align}
the exponential term becomes
\begin{align}
-\frac{e^2}{2}<\int A\int A>_{\rm exp}=
\frac{e^2}{2\pi^2}\bigl[\,\delta\cot\delta-1\,\bigr]K_0(m_\gamma t)
+R(m_\gamma t)
\label{k0-decomposition}
\end{align}
where
\begin{align*}
R(m_\gamma t)&
=-\frac{e^2}{4\pi^2}\int_0^\infty d\eta\;
e^{-m_\gamma t\cosh\eta}
\biggl[\frac{\sinh^2\eta}{\cosh^2\eta-\sin^2\delta}
+\tanh^2\eta
\\&
-2\cot\delta\,\tanh\eta\,\arctan\!\bigl(\tanh\eta\,\tan\delta\bigr)
-2+2\,\delta\cot\delta\biggr]
\end{align*}
has an integrand that vanishes exponentially at large $\eta$, so that
$R$ is finite as $m_\gamma t\to0$ and exponentially small as
$m_\gamma t\to\infty$. Thus $K_0(m_\gamma t)$ is the exact crossover
function of the problem. Its two asymptotic regimes,
\begin{align}
K_0(w)=-\ln\frac{w}{2}-\gamma_E+O\!\left(w^2\ln w\right),
\qquad
K_0(w)=\sqrt{\frac{\pi}{2w}}\,e^{-w}\left(1+O(1/w)\right),
\label{k0-asymptotics}
\end{align}
interpolate between the infraparticle logarithm for $t<<1/m_\gamma$ and the
exponential switching-off of the anomalous power beyond $t\sim1/m_\gamma$.
In particular, in the window $1/m<< t<<1/m_\gamma$,
\begin{align}
\frac{e^2}{2\pi^2}\bigl[\delta\cot\delta-1\bigr]K_0(m_\gamma t)
\longrightarrow
\frac{e^2}{2\pi^2}
\left[\gamma_E+\ln\frac{m_\gamma t}{2}\right]
\bigl[1-\delta\cot\delta\bigr]
\label{small-mgamma-limit}
\end{align}
with $m_\gamma t$ rather than$\mu t$, inside the logarithm.  The exponential piece
is ultraviolet finite and it should therefore not depend on $\mu$, and, indeed, the $\mu$-dependence cancels.

As a consistency check, let us confirm that we recover the result of section \ref{wilson lines} in the 
limit where we put the photon mass to zero. 
Adding the frozen term (\ref{frozen-result}) to the small-$m_\gamma$ limit
(\ref{small-mgamma-limit}) of the exponential term, we get
\begin{align}\nonumber
&\lim_{m_\gamma\to 0}~-\frac{e^2}{2}<\int A\int A>=
\\&
\frac{e^2}{4\pi^2}\bigl[1-\delta\cot\delta\bigr]
\left[\frac{1}{\epsilon}-\gamma_E+\ln4\pi
+\ln\frac{\mu^2}{m_\gamma^2}
+2\gamma_E+\ln\frac{m_\gamma^2t^2}{4}\right]
\nonumber \\ &=\frac{e^2}{4\pi^2}
\left[\frac{1}{\epsilon}+\gamma_E+\ln4\pi
+\ln\frac{\mu^2t^2}{4}\right]
\bigl[\,1-\delta\cot\delta\,\bigr]
\label{consistency}
\end{align}
The logarithms of $m_\gamma$ cancel identically between the two pieces. The final expression 
is similar to the massless-photon result of section \ref{wilson lines}.  It contains the pole
$1/\epsilon$ with coefficient $\frac{e^2}{4\pi^2}(1-\delta\cot\delta)$
paired with $\ln(\mu^2 t^2/4)$, whose $t$-dependence is the anomalous
dimension
$
\gamma=\frac{e^2}{2\pi^2}\bigl(\delta\cot\delta-1\bigr)
$.   

Its angle $\delta$ is now the angle between the dressing velocity and the
displacement $x$.  In the semi-classical limit the amplitude is dominated
by the classical trajectory, $p^\mu = m x^\mu/t$, so $\delta$ is also the
Euclidean angle between the dressing velocity and the momentum, and the
identification $\zeta=i\delta$ gives $\zeta\coth\zeta=\delta\cot\delta$,
reproducing the exponent of section \ref{wilson lines}.  For real
Euclidean angles $\gamma(\delta)<0$ and the anomalous power
$t^{-\gamma(\delta)}$ is an enhancement; the inverse Wick rotation
$\delta\to-i\zeta$ turns it into the real-time suppression
$t^{-\gamma(\zeta)}$ with $\gamma(\zeta)\geq0$.

To proceed, we perform a numerical computation of the crossover function,\\ $\exp\left(-\frac{e^2}{2}<\int A\int A>\right)$.  The result is displayed in figure \ref{fig:crossover}. There we can see that the transition region between the asymptotic regimes is relatively narrow.

\section{Concluding remarks}

\label{conclusion}

In this paper we have presented a perturbative exploration of the effect of the dressing that is needed in order to make operators that create an electric charge gauge invariant, and in order to make their 2 point and higher point correlation functions gauge invariant. We specifically studied two point correlation functions of Wilson-line-dressed complex scalar field operators in the context of scalar quantum electrodynamics. Our main results have already been well outlined and discussed in section \ref{introduction}. 

Here, we note that our quantum electrodynamics discussion has a well-motivated and closely related narrative in quantum gravity, both the perturbative and nonperturbative versions.  From the nonperturbative perspective it has long been said that quantum gravity can have no local observables and seemingly local observations are only defined relationally with respect to standard clocks and rulers \cite{dewitt}. The use of gravitational dressing to define observables in both nonperturbative and perturbative gravity has seen some development recently \cite{Gid1,Gid2,Gid3,Choi:2017bna,frob,massless}.  

We have little to add to this beyond the statement that the need for dressing seems even more pressing in the context of general coordinate invariance.  Once there, it requires the attachment of entities carrying long-ranged fields to local operators.  The ensuing infrared properties of those long-ranged fields, which in perturbative quantum gravity about flat space, take on a very similar structure to those of QED, should have consequences for gravity, analogous to those that we have outlined for quantum electrodynamics. However, the details, for example an analog of our St\"uckelberg construction of a gauge invariant photon mass, are underdeveloped for gravity, and they do pose interesting fundamental questions.

\acknowledgments
This work is supported in part by the Natural Sciences and Engineering Research Council of Canada.   The authors further acknowledge the use of Anthropic Fable AI to check the accuracy of computations and references.

\end{document}